\documentclass[journal]{IEEEtran}
\usepackage{amsfonts}
\IEEEoverridecommandlockouts

\ifCLASSINFOpdf
\else
\fi
\usepackage{multirow}
\usepackage{cite}
\usepackage{color}
\usepackage{graphicx}
\usepackage{amssymb }
\usepackage{algorithm}
\usepackage{algorithmic}
\usepackage{booktabs}
\usepackage{amsmath,bm}
\usepackage{colortbl}
\usepackage{makecell}
\usepackage[colorlinks]{hyperref}

\newtheorem{theorem}{Theorem}

\newtheorem{lemma}{Lemma}
\newtheorem{remark}{Remark}

\usepackage[percent]{overpic}
\begin{document}

\title{Performance Analysis of Single-Antenna Fluid Antenna Systems via Extreme Value Theory}

\author{Rui Xu,
            Yinghui Ye,~\IEEEmembership{Member,~IEEE},
            Xiaoli Chu,~\IEEEmembership{Senior Member,~IEEE},
            Guangyue Lu,\\
            Kai-Kit Wong,~\IEEEmembership{Fellow,~IEEE}, and
            Chan-Byoung Chae,~\IEEEmembership{Fellow,~IEEE}
            \vspace{-8mm}

\thanks{This work was supported by the National Natural Science Foundation of China under Grants  62571430 and 62471388.}
\thanks{R. Xu, Y. Ye, and G. Lu are with the Shaanxi Key Laboratory of
Information Communication Network and Security, Xi'an University of Posts
\& Telecommunications, China (e-mails: $\rm dhlscxr@126.com$; $\rm connectyyh@126.com$; $\rm tonylugy@163.com$). }
\thanks{X. Chu is with the School of Electrical and Electronic Engineering, University of Sheffield, Sheffield, U.K. (e-mail: $\rm x.chu@sheffield.ac.uk$).}
\thanks{K. K. Wong is with the Department of Electronic and Electrical Engineering, University College London, WC1E 7JE, London, United Kingdom (e-mail: $\rm{kai\text{-}kit.wong}@ucl.ac.uk$). K. K. Wong is also affiliated with Yonsei Frontier Laboratory, Yonsei University, Seoul, 03722, Republic of Korea.}
\thanks{C.-B. Chae is with the School of Integrated Technology, Yonsei University, Seoul, 03722, Republic of Korea (e-mail: $\rm cbchae@yonsei.ac.kr$).}}

\markboth{}
{Shi\MakeLowercase{\textit{et al.}}:}
\maketitle
\begin{abstract}
In a single-antenna fluid antenna system (FAS), the transceiver dynamically selects the antenna port with the strongest instantaneous channel to enhance link reliability. However, deriving accurate yet tractable performance expressions under the fully correlated fading channel remains challenging, primarily due to the absence of a closed-form distribution for the FAS channel. To address this gap, this paper develops a novel performance evaluation framework for FAS under fully correlated Rayleigh fading by modeling the FAS channel using extreme value distributions (EVDs).
We first justify the suitability of EVDs and model the FAS channel as a Gumbel distribution, whose parameters, estimated via the maximum likelihood (ML) criterion, are expressed as functions of the number of ports and antenna aperture size.
Closed-form approximate expressions for the outage probability (OP) and the ergodic capacity (EC) are then derived. Simulation results show that the Gumbel distribution provides simple yet sufficiently accurate expressions for EC, although slight deviations in OP accur in the low-probability region of practical interest. To further improve accuracy, the FAS channel is modeled using the generalized extreme value (GEV) distribution, yielding closed-form expressions for OP and EC based on ML-estimated parameters. Simulation results confirm that the GEV distribution provides superior accuracy compared to the Gumbel, particularly for OP, while both EVD-based approaches offer computationally efficient and analytically tractable tools for evaluating FAS performance under the fully correlated fading channel.
\end{abstract}


\IEEEpeerreviewmaketitle

\vspace{-2mm}
\section{Introduction}
\IEEEPARstart{F}{luid} antenna system (FAS) has emerged as a promising physical-layer technology for 6G wireless networks, capable of significantly enhancing transmission reliability by exploiting the full extent of spatial diversity within a confined region \cite{10753482}. In FAS, a large number of antenna ports are densely deployed, and the port with the
strongest instantaneous channel, termed the FAS channel, is dynamically selected for signal transmission or reception. This adaptive port selection enables substantial performance gains without requiring multiple radio frequency (RF) chains.
However, the dense spatial deployment of ports inevitably leads to spatial correlation among the channel envelopes, making it analytically intractable to characterize the distribution of the FAS channel. Consequently, deriving accurate yet tractable performance expressions for FAS remains a significant theoretical challenge, which motivates extensive research efforts \cite{10103838, 10130117,11023237, 10965723,10924151,10253941,10678877,11230883,9264694, wong2022closed,10623405}\footnote{Beyond performance evaluation, extensive research efforts have addressed other fundamental challenges in FAS, encompassing channel estimation \cite{10807122,10751774,11180048}, beamforming design \cite{11184546,11219181,10508198}, and performance optimization \cite{11196915,11098708,10794752,10772590}.
For example, a successive Bayesian reconstructor for FAS channel estimation was proposed in \cite{10807122}, and a novel estimation technique based on latent-domain representation was proposed in \cite{11180048} to identify the required training overhead.
Moreover, beamforming design for FAS was investigated in \cite{11184546,11219181}, and a deep learning-based method was introduced for beamforming design in \cite{10508198} to improve computational efficiency.
From a performance optimization perspective, various works have focused on optimizing key metrics such as bit error rate, capacity, and sum-rate, e.g., \cite{11196915,11098708,10794752,10772590}.}.

To enable accurate performance evaluation, a precise correlated channel model that reflects the realistic spatial correlation structure in FAS is essential.
Towards this end, \cite{10103838} introduced the fully correlated channel model where a correlation matrix based on Jakes' model \cite{Stuber2002} characterizes the expected spatial correlation between the channels at any two ports, and the channel at each port is modeled as a linear combination of those at other ports. Based on the eigenvalue decomposition of the correlation matrix, the cumulative distribution function (CDF) expression of the FAS channel involves multiple nested integrals, making performance evaluation of FAS intractable. To strike a balance between tractability and accuracy, extensive research has been conducted mainly from three aspects: reducing or eliminating multi-fold integrals  \cite{10103838, 10130117,11023237,10965723,10924151}, modeling the channel distribution using copula theory \cite{10253941,10678877,11230883}, and simplifying the spatial correlation structure \cite{9264694, wong2022closed,10623405}.

\emph{Reducing or eliminating multi-fold integrals:} The authors in \cite{10103838} proposed a two-stage approximation in which the first-stage approximation reduces the number of multi-fold integrals in the outage probability (OP) expression by considering only the few dominant eigenvalues, while the second-stage approximation expresses the OP in the form of a single integral raised to a power. However, the approximate results align well with simulation results only when the number of ports is small. Although \cite{10130117} further approximated the OP of FAS in a finite series expression based on the series expansion approach in \cite{wiegand2020new}, the derived expression is complicated and its validity was demonstrated through computer simulations only for cases with a small number of ports. Furthermore, in \cite{11023237}, the outage expression of FAS was derived in the form of an infinite series by approximating the covariance matrix with an exponential correlation matrix following the approach in \cite{1230126}, while the computational complexity remains high despite the reduced number of summation terms.
Given that characterizing the lower tail of the FAS channel for OP evaluation is analytically challenging, \cite{10965723} instead analyzed the upper tail of the signal-to-noise ratio (SNR) in a multi-dimensional space for continuous FAS. By employing the level crossing probability for one-dimensional case and the expected Euler characteristic for arbitrary high-dimensional cases, this approach avoids multi-fold integrals and yields a closed-form expression for the high-SNR probability. Based on the derived diversity gain, the FAS channel in \cite{10924151} was approximated as the maximum of multiple independent channels, and a closed-form expression for the OP was derived.
Even though the above methods reduce or eliminate the multi-fold integrals for OP evaluation under the fully correlated channel model, the resulting expressions remain either inaccurate or still complicated.

\emph{Modeling channel distribution using copula theory:} Considering that, in copula theory, the multivariate distribution of correlated random variables (RVs) can be generated based on the marginal distributions, and that the dependency structure can be modeled through a copula function, copula theory was employed to model the FAS channel distribution based on different types of copula. In \cite{10253941}, a closed-form OP expression was derived under arbitrary fading channels using the family of Archimedean copulas, but this type of copula cannot capture the complex dependence structure with a fixed dependence parameter. To address this issue, the Gaussian copula, a type of elliptical copula, has attracted considerable attention since it incorporates a covariance matrix to capture the dependence among multivariate normal RVs. Based on this, the covariance matrix was approximated using the channel coefficient correlation matrix from Jakes' model in \cite{10678877}, and an approximate OP was derived. Accordingly, the Gaussian copula links the FAS performance to the number of ports and the size of antenna, enabling more accurate modeling under different parameters, while the derived OP only aligns with computer simulations in the dense port deployment. Considering that the covariance matrix in the Gaussian copula reflects the dependence among multivariate normal RVs transformed from the correlated channel envelopes, the authors in \cite{11230883} proposed to approximate the covariance matrix using the envelope-level correlation matrix, and verified its superior accuracy particularly in the sparse port deployment and low-outage regime. However, it should be noted that in the Gaussian copula, the multivariate normal CDF is computed using the MATLAB function {\tt mvncdf}, which involves high-dimensional integration. As the number of ports increases, the computational complexity grows rapidly, rendering this approach impractical for large-scale port scenarios.

\emph{Simplifying the spatial correlation structure:} Under the fully correlated channel model, the realistic spatial correlation structure complicates the channel model and makes the performance evaluation more challenging. Many studies simplified the correlation structure and developed simplified correlated channel models to facilitate performance characterization. In \cite{9264694}, a simple correlation model was considered, where the first port was treated as a reference port, and only the correlation between the channel at another port and that at the reference port was characterized using Jakes' model. Under this model, the authors obtained the exact and approximate OPs in integral-form and closed-form expressions, respectively, but this model fails to capture the channels correlation at any two ports, potentially leading to performance overestimation.

Moreover, a common correlation coefficient was employed in \cite{wong2022closed} to describe the correlation between the channels at any two ports, and an integral-form OP expression was derived. Considering that a common average-correlation parameter may not precisely characterize the correlation, \cite{10623405} proposed a block-diagonal correlation model, where the spatial correlation of the channels remains approximately constant within each block (a subset of ports) while different blocks are assumed independent. Based on this, the integral-form and closed-form OP expressions were derived for exact and approximate performance, respectively. All of the above models simplify the realistic spatial correlation structure, at the cost of reduced accuracy in characterizing the practical performance of FAS.

From the above literature review, it is clear that the existing performance evaluations for single-antenna FAS are yet to strike a proper balance between accuracy and tractability. This is because they fail to accurately capture the FAS channel distribution with satisfactory precision while maintaining low complexity. In this paper, we approximate the analytically intractable FAS channel distribution using a known parametric distribution via distribution fitting techniques. This technique has been widely used in wireless communications to model channel distributions (e.g., see \cite{7336572,6059452,8809411,11106228}). However, two key issues need to be addressed when using this technique: i) Which distributions can more accurately fit the distribution of the FAS channel while maintaining tractability? ii) How do the parameters of the fitted distribution correspond to the FAS parameters?

To address the above issues, this paper develops an extreme value distribution\footnote{The EVD consists of the Gumbel, the {\text{Fr\'echet}}, and the Weibull distributions, all of which can be represented as members of a single family of the generalized EVD \cite{leadbetter2012extremes}.} (EVD)-based fitting framework\footnote{The best channel selected from multiple correlated candidates closely parallels the concept of extreme events, which naturally motivates us to investigate the feasibility of employing the EVD to fit the distribution of the FAS channel. This method has been widely applied in wireless channel modeling  \cite{9508780, 9462379} and performance evaluation \cite{8809213,1710338}.} that enables accurate and tractable performance evaluation of FAS under the fully correlated Rayleigh fading channel. The main contributions are summarized as follows:
\begin{itemize}
\item We first analyze the rationale for fitting the FAS channel distribution using the same EVD of corresponding independent and identically distributed (i.i.d.) RV sequence with the modified normalizing parameters, from the perspective of weak dependence and realistic data modeling.
\item Given that the limiting distribution of the maximum of i.i.d. Rayleigh RVs is the Gumbel distribution, we first fit the FAS channel distribution with the Gumbel distribution using the distribution fitting technique based on the maximum likelihood (ML) criterion. The scale and location parameters of the Gumbel distribution, corresponding to the normalizing parameters, are obtained as functions of the number of ports and antenna size. Using the fitted Gumbel distribution, closed-form expressions for OP and EC are derived to evaluate the FAS performance.
Simulation results show that the fitted Gumbel distribution closely matches the FAS channel distribution in general, while some deviations are present in the high- and ultra-low-probability regions. As a result, a relatively accurate approximation of EC can be achieved, whereas slight deviations are observed between the approximate OP and the Monte-Carlo simulation results, particularly in the low-probability region of practical interest.
Nevertheless, the Gumbel distribution is simple and accurate enough for EC, but the more accurate EVD is generally needed for OP because the low-probability region is essential and that's where the deviation occurs.\footnote{The Gumbel distribution, as a theoretically grounded model, provides a reference for understanding the impact of a finite number of ports and channel correlation, motivating the use of a more flexible EVD to model the FAS channel.}

\item To improve the fitting accuracy, we fit the FAS channel distribution using a more flexible GEV distribution, which incorporates an additional shape parameter to adjust the EVD type and has more flexible skewness and kurtosis to control the symmetry and tail behavior.
Based on this, we obtain the scale, location and shape parameters of the GEV distribution as functions of the number of ports and antenna size, respectively, and derive closed-form expressions for OP and EC of FAS.
Simulation results show that the GEV distribution provides a superior fit to the FAS channel, resulting in more accurate OP evaluation and improved EC accuracy.
Moreover, the GEV distribution achieves accurate approximations of OP and EC with much lower computational complexity compared to existing methods.
\end{itemize}

The remainder of this paper is organized as follows: Section \ref{sec:background} presents the system model of FAS and the classical extreme value theory (EVT). Section \ref{sec:evd} then discusses the rationale for modeling the FAS channel using the EVD from the perspective of weak dependence and realistic data modeling. Sections \ref{sec:gumbel} and \ref{sec:gev} evaluate the FAS performance using the Gumbel and GEV distributions, respectively, including channel distribution fitting, EVD parameter fitting, approximate OP and EC derivations, and simulation results. Finally, we conclude the paper in Section \ref{sec:conclude}.

\section{Background}\label{sec:background}
\subsection{System Model of FAS}\label{ssec:model}
In this paper, we consider a point-to-point FAS, in which a single fixed-antenna transmitter sends information to a receiver equipped with a fluid antenna.\footnote{Note that the term `fluid' here is to highlight the flexible nature, rather than refer to the use of fluidic materials.} The fluid antenna is connected to one RF-chain and there are ${N}$ preset antenna ports evenly distributed along a straight line of length ${W}\lambda$, where $\lambda$ is the carrier wavelength and $W$ denotes the antenna size. Therefore, the distance between the $i$-th and the $j$-th port, $i,j \in {\cal N} = \left\{ {1, \dots ,N} \right\}$, is given by ${d_{i,j}} = \frac{{\left| {i - j} \right|}}{{N - 1}}W\lambda$. Based on Jakes' model, we define the spatial correlation matrix as ${\bf{J}} \in {{\mathbb{C}}^{N \times N}}$, where its $\left( {i,j} \right)$-th element can be expressed as ${{\bf{J}}_{i,j}} = {J_0}\left( {{{2\pi W\left( {\left| {i-j} \right|} \right)} \mathord{\left/ {\vphantom {{2\pi W\left( {\left| {\bar n - \tilde n} \right|} \right)} {\left( {N - 1} \right)}}} \right. \kern-\nulldelimiterspace} {\left( {N - 1} \right)}}} \right)$, and ${J_0}\left(  \cdot  \right)$ is the zero-order Bessel function of the first kind.

 As explained in \cite{10103838}, the fully correlated Rayleigh fading channel ${\bf{h}}= {\left[ {{h_1}, \dots {h_N}} \right]^T}\in {{\mathbb{C}}^{N \times 1}}$ at the $N$ ports with spatial correlation properties ${\bf{J}}$ can be expressed in the form of ${\bf{U}}{{\bf{\Lambda }}^{\frac{1}{2}}}{\bf{z}}$, where ${\bf{z}} = {\left[ {{z_1}, \dots {z_N}} \right]^T}$, whose elements are i.i.d.~circularly symmetric complex Gaussian RVs with zero mean and unit variance, ${\bf{\Lambda}} ={\rm{diag}}\left( {{\lambda _1}, \dots ,{\lambda _N}} \right)$ denotes the diagonal eigenvalue matrix of ${\bf{J}}$ with ${\lambda _1} \ge \cdots \ge {\lambda _N}$ and can be obtained by ${\bf{J}} = {\bf{U\Lambda }}{{\bf{U}}^{\mathcal{H}}}$ with the conjugate transpose $\mathcal{H}$ and $\bf{U}$ is the eigenvector matrix. By doing so, ${\bf{h}} \sim \mathcal{CN}\left( {0,{\bf{J}}} \right)$ and each element ${h_i}$ follows a Rayleigh distribution with scale parameter $\sigma  = {1 \mathord{\left/ {\vphantom {1 {\sqrt 2 }}} \right. \kern-\nulldelimiterspace} {\sqrt 2 }}$.

Similar to the prior efforts \cite{10103838, 10678877,9264694, 10623405}, the receiver activates the antenna port with the maximum channel envelope to receive the signal, i.e.,
\begin{align}\label{7A}
\left| {{h_{{\rm{FAS}}}}} \right| = \max \left\{ {\left| {{h_{\rm{1}}}} \right|, \left| {{h_{\rm{2}}}} \right|, \dots ,\left| {{h_N}}\right|}\right\}.
\end{align}
The instantaneous received SNR is given by ${\left| {{h_{{\rm{FAS}}}}} \right|^2}\bar \gamma $, where $\bar \gamma  = {P \mathord{\left/
 {\vphantom {P {{N_0}}}} \right. \kern-\nulldelimiterspace} {{N_0}}}$ denotes the average transmit SNR with the transmit power $P$ and the noise power ${N_0}$. As such, the OP and EC of the considered FAS are defined as
\begin{align}
{P_{{\rm{out}}}} &= \Pr \left\{ {\left| {{h_{{\rm{FAS}}}}} \right| \le \hat \gamma } \right\},\label{8A}\\
\bar C &= \mathbb{E}\left\{ {\ln \left( {1 + {{\left| {{h_{{\rm{FAS}}}}} \right|}^2}\bar \gamma } \right)} \right\}{\rm{\;\;\;}}{{{{{\rm{nats}}} \mathord{\left/
 {\vphantom {{{\rm{nats}}} {\rm{s}}}} \right.
 \kern-\nulldelimiterspace} {\rm{s}}}} \mathord{\left/
 {\vphantom {{{{{\rm{nats}}} \mathord{\left/
 {\vphantom {{{\rm{nats}}} {\rm{s}}}} \right.
 \kern-\nulldelimiterspace} {\rm{s}}}} {{\rm{Hz}}}}} \right.
 \kern-\nulldelimiterspace} {{\rm{Hz}}}},\label{9A}
\end{align}
where $\hat \gamma  = \sqrt {{{{\gamma _{{\rm{th}}}}} \mathord{\left/ {\vphantom {{{\gamma _{{\rm{th}}}}} {\bar \gamma }}} \right. \kern-\nulldelimiterspace} {\bar \gamma }}}$, and ${{\gamma _{{\rm{th}}}}}$ is the decoding SNR threshold.

\begin{remark}
The challenge in evaluating the OP and EC lies in deriving the distribution of $\left| {{h_{{\rm{FAS}}}}} \right|$, which is the maximum of correlated and identically distributed (c.i.d.) RVs. However, the existing approaches, including integration methods and copula function modeling, would not be able to derive an accurate and tractable expression. Since the concept of deriving the distribution of the maximum of multiple RVs is similar to the concept of EVT, we fit the distribution of $\left| {{h_{{\rm{FAS}}}}} \right|$ using the EVD to evaluate the FAS performance. Although the classical EVT describes the limiting law of the extreme value of the i.i.d.~RV sequence, the distributions in EVT can fit the extreme value of the c.i.d. RV sequence well under certain conditions and this will be discussed in Section \ref{sec:evd}. To aid understanding, we provide a brief introduction to EVT in Section \ref{ssec:evt}, and then explain the rationale for modeling the distribution of the maximum of c.i.d.~RV sequence using the EVD in Section \ref{sec:evd}.
\end{remark}

\subsection{EVT}\label{ssec:evt}
EVT is a powerful tool for characterizing the probabilistic distribution of extreme events occurring with low probability. It provides an elegant statistical tool for studying the asymptotic distributions of the maximum or minimum of a set of RVs \cite{leadbetter2012extremes, de2006extreme}. Let $\left\{ {{X_1}, \dots ,{X_n}} \right\}$ be an i.i.d.~RV sequence, with the CDF of each being $F\left( x \right)$, and ${M_n} = \max \left\{ {{X_1}, \dots ,{X_n}} \right\}$. Suppose that there exist two sequences of real numbers $\left\{ {{a_n} > 0} \right\}$ and $\left\{ {{b_n} \in \mathbb{R}} \right\}$, referred to as the normalizing parameters, such that the following limits converge to a nondegenerate distribution function, given by
\begin{align}\label{1A}
\mathop {\lim }\limits_{n \to \infty } \Pr \left( {\frac{{{M_n} - {b_n}}}{{{a_n}}} \le y} \right) = G\left( y \right).
 \end{align}
By the Fisher-Tippett-Gnedenko theorem, the limiting function $G$ is the CDF of a distribution belonging to either the Gumbel, {\text{Fr\'echet}}, or Weibull distribution, defined as
\begin{align}
{G_1}\left( y \right)\! &=\!\exp \left( { - {e^{ - y}}} \right),y \in \mathbb{R},{\rm{  }}\left( {{\text{Gumbel}}} \right),\label{2A}\\
{G_2}\left( {y;\alpha } \right)\!& =\! \left\{ {\begin{array}{*{20}{c}}
{0,\;\;\;\;\;\;\;\;\;\;\;\;\;\;\;\;\;y \le 0,}\\
{\exp \left( { - {y^{ - \alpha }}} \right),y > 0,}
\end{array}} \right.\!\!{\rm{ }}\alpha  > 0,{\rm{  }}\left( {\text{Fr\'echet}} \right),\label{3A}\\
{G_3}\left( {y;\alpha } \right) \!&=\! \left\{ {\begin{array}{*{20}{c}}
{\exp \left( { - {{\left( { - y} \right)}^\alpha }} \right),y \le 0,}\\
{1,\;\;\;\;\;\;\;\;\;\;\;\;\;\;\;\;\;\;\;\;y > 0,}
\end{array}} \right.\!\!{\rm{ }}\alpha  > 0.{\rm{  }}\left( {{\text{Weibull}}} \right).\label{4A}
\end{align}
The above three types of distributions may all be represented as members of a single family of GEV distributions with CDF
\begin{align}\label{5A}
G\left( {y;\xi } \right) = \exp \left( { - {{\left( {1 + \xi y} \right)}^{ - {1 \mathord{\left/{\vphantom {1 \xi }} \right.\kern-\nulldelimiterspace} \xi }}}} \right),1 + \xi y > 0,
\end{align}
where the shape parameter, $\xi$, can be any real number. For $\xi  > 0,\alpha  = {1 \mathord{\left/ {\vphantom {1 \xi }} \right.\kern-\nulldelimiterspace} \xi }$, $G\left( {y;\xi } \right)$ denotes the \text{Fr\'echet} distribution; for $\xi  = 0$, $G\left( {y;\xi } \right)$ denotes the \text{Gumbel} distribution; for $\xi  < 0,\alpha  =  - {1 \mathord{\left/ {\vphantom {1 \xi }} \right. \kern-\nulldelimiterspace} \xi }$, $G\left( {y;\xi } \right)$ denotes the \text{Weibull} distribution.

The distributions shown in (\ref{2A})--(\ref{5A}) are the standard EVDs, which can be generalized by introducing the location parameter $\mu$ and scale parameter $\sigma$. The resulting distributions can be expressed as ${G_1}\left( {\frac{{y - \mu }}{\sigma }} \right)$, ${G_2}\left( {\frac{{y - \mu }}{\sigma };\alpha } \right)$, ${G_3}\left( {\frac{{y - \mu }}{\sigma };\alpha } \right)$ and $G\left( {\frac{{y - \mu }}{\sigma };\xi } \right)$, respectively. According to \eqref{1A}, we can see that $\mathop {\lim }\limits_{n \to \infty } \Pr \left( {{M_n} \le {a_n}y + {b_n}} \right) = G\left( y \right)$. Letting $x = {a_n}y + {b_n}$, $\mathop {\lim }\limits_{n \to \infty } \Pr \left( {{M_n} \le x} \right) = G\left( {{{\left( {x - {b_n}} \right)} \mathord{\left/ {\vphantom {{\left( {x - {b_n}} \right)} {{a_n}}}} \right. \kern-\nulldelimiterspace} {{a_n}}}} \right)$ can be derived.\footnote{It should be emphasized that the normalizing parameters $a_n$ and $b_n$ are not unique. Different choices may lead to different limit distributions, which are related to each other as established in the theorem in \cite{leadbetter2012extremes} which is given here. Assume that there exist constants $a_n$, $b_n$, $c_n$, and $d_n$, such that $\mathop {\lim }\limits_{n \to \infty } \Pr \left( {{M_n} \le {a_n}y + {b_n}} \right) = G\left( y \right)$ and $\mathop {\lim }\limits_{n \to \infty } \Pr \left( {{M_n} \le {c_n}y + {d_n}} \right) = {G^ * }\left( y \right)$. There exist constants $A$ and $B$ such that ${G^ * }\left( y \right) = G\left( {Ay + B} \right)$ holds, where $G\left( y \right)$ and ${G^ * }\left( y \right)$ have the same distribution type and $A = {{{c_n}} \mathord{\left/ {\vphantom {{{c_n}} {{a_n},}}} \right. \kern-\nulldelimiterspace} {{a_n}}}$, $B = {{\left( {{d_n} - {b_n}} \right)} \mathord{\left/
 {\vphantom {{\left( {{d_n} - {b_n}} \right)} {{a_n}}}} \right. \kern-\nulldelimiterspace} {{a_n}}}$. We only consider the normalizing parameters, ensuring that the normalized maximum converges to a standard EVD.} We note that the normalizing parameters $a_n$ and $b_n$ correspond to the scale parameter $\sigma$ and the location parameter $\mu$ of the generalized EVD, respectively. Accordingly, the normalization of the maxima can be reflected in the scale and location parameters of the EVD, and the maxima of the sequence can be directly fitted by the EVD.

When modeling realistic data with the EVD, the procedure is generally as follows \cite{majumdar2014extreme}: First, let the observation sequence be $\left\{ {{x_1}, \dots ,{x_n}} \right\}$, which can be evenly partitioned into $k$ blocks of length $m$. Second, the maximum value is extracted from each block, yielding the block maxima $\left\{ {{y_1}, \dots ,{y_k}} \right\}$. According to the Fisher-Tippett-Gnedenko theorem, when $m$ is sufficiently large, $\left\{ {{y_1}, \dots ,{y_k}} \right\}$ can be approximated as i.i.d.~samples from the EVD. Third, the distribution of the extreme value can be obtained by modeling $\left\{ {{y_1}, \dots ,{y_k}} \right\}$ using the EVD. Although the classical framework of EVT typically assumes that the underlying RVs are i.i.d., many realistic datasets exhibit dependence, such as time series. Nevertheless, under certain weak-dependence conditions, the dependence among the block maxima $\left\{ {{y_1}, \dots ,{y_k}} \right\}$ becomes asymptotically negligible. This enables the EVD to be effectively applied to fit the distribution of extreme value of dependent sequence.

\section{The Rationale for Fitting the FAS Channel Distribution using the EVD}\label{sec:evd}
In this section, we provide the rationale for fitting the FAS channel distribution using the EVD from the perspective of weak dependence and realistic data modeling.

\subsection{Weak Dependence}\label{ssec:weak}
It has been pointed out in \cite[Chapter 3]{leadbetter2012extremes} that the classical EVT results remain true if the condition of i.i.d. RVs is replaced by the requirement that they form a stationary sequence satisfying a very weak dependence restriction. For a better understanding, we reproduce some related conclusions from \cite{leadbetter2012extremes} as \emph{Lemmas~1}$-$\emph{3} as follows.

Before presenting these lemmas, we find it useful to first introduce the following definitions:
\begin{enumerate}
\item We denote $\left\{ {{X_1},{X_2}, \dots ,{X_n}} \right\}$ as an i.i.d.~RV sequence and also $\{ {{{\tilde X}_1},{{\tilde X}_2}, \dots ,{{\tilde X}_n}}\}$ a c.i.d.~RV sequence, where each $\tilde X_i$ has the same distribution as $X_i$, $i \!=\! 1,\dots, n$.
\item Define ${M_n}\!\triangleq\! \max \left\{ \!{{X_1}\!,\!\dots \!,\!{X_n}}\! \right\}$, ${{\tilde M}_n} \!\triangleq\!\max\{\! {{{\tilde X}_1}\!,\!\dots \!,\!{{\tilde X}_n}} \!\}$.
\item The joint distribution function of $\{{ \tilde X_{{i_1}}}, \dots ,{\tilde X_{{i_n}}}\}$ is denoted by
$${F_{{i_1}, \dots ,{i_n}}}\!\left( \!{{{ x}_1}, \!\dots\! ,{{ x}_n}} \!\right)\triangleq\Pr \!\left\{ \! {{{\tilde X}_{{i_1}}} \le {{ x}_1},} \right. \!\dots\! ,\left. {{{\tilde X}_{{i_n}}} \le {{x}_n}} \!\right\},$$
and for brevity, ${F_{{i_1}, \dots ,{i_n}}}\!\left(u \right) \triangleq {F_{{i_1}, \dots ,{i_n}}}\!\left( {u, \dots ,u}\right)$ for each $n$, ${i_1, \dots ,{i_n}}$ and $u$.
\item The normalizing parameters of ${M_n}$ are denoted by ${a_n}$ and ${b_n}$, and ${u_n} = {a_n}y + {b_n}$.
\end{enumerate}

\begin{lemma}\label{lemma1}
 Definition of the weak dependence $D(u_n)$ condition. The sequence $\{ \tilde X_1, \dots , \tilde X_n\}$ is considered satisfying the $D(u_n)$ condition if for any integers $n$, $l_n$, ${i_1}, \dots ,{i_p},{j_1}, \dots ,{j_{p'}}$ such that
\begin{align}\label{13A}
1 \le {i_1} <  \cdots  < {i_p} < {j_1} <  \cdots  < {j_{p'}} \le n,\;{j_1} - {i_p} \ge l_n,
\end{align}
we have
\begin{align}\label{14A}
\left| {{F_{{i_1}, \dots ,{i_p},{j_1}, \dots ,{j_{p'}}}}\left( {{u_n}} \right)-{F_{{i_1}, \dots ,{i_p}}}\left({{u_n}}\right){F_{{j_1}, \dots ,{j_{p'}}}}\left({{u_n}} \!\right)} \right| < {\alpha _{n,l_n}},
\end{align}
where ${\alpha _{n,l_n}} \to 0$ as $n \to\infty $ for some sequence ${l_n} = o\left( n \right)$, i.e., ${{{l_n}} \mathord{\left/ {\vphantom {{{l_n}} n}} \right. \kern-\nulldelimiterspace} n} \to 0$.
\end{lemma}

\begin{lemma}\label{lemma2}
Let $\{ {{{\tilde X}_1}, \dots ,{{\tilde X}_n}} \}$ be a stationary sequence, and suppose that $\Pr \left\{ {{{\left( {{\tilde M_n} - {b_n}} \right)} \mathord{\left/{\vphantom {{\left( {{\tilde M_n} - {b_n}} \right)} {{a_n}}}} \right. \kern-\nulldelimiterspace} {{a_n}}} \le y} \right\}$ converges to a nondegenerate distribution function $G\left( y \right)$. If $D\left( {{u_n}} \right)$ condition is satisfied for ${u_n} = {a_n}y + {b_n}$ for each real $y$, then $G\left( y \right)$ has one of the three extreme value forms, as shown in \eqref{2A}--\eqref{4A}.
\end{lemma}

\begin{lemma}\label{lemma3}
Let the stationary sequence $\{ {{{\tilde X}_1}, \dots ,{{\tilde X}_n}}\}$ have extremal index $\theta  \in \left( {0,1} \right]$, which characterizes the degree of clustering of extreme values in a stationary sequence. Then ${{{\hat M}_n}}$ has a nondegenerate limiting distribution if and only if ${M_n}$ does, and both distributions are of the same type. Furthermore, the same normalization may be applied, or the normalizing parameters ${a_n}$ and ${b_n}$ may be adjusted to yield exactly the same limiting distribution function. Specifically, if $\Pr \left\{ {{{\left( {{M_n} - {b_n}} \right)} \mathord{\left/
 {\vphantom {{\left( {{M_n} - {b_n}} \right)} {{a_n} \le y}}} \right.
 \kern-\nulldelimiterspace} {{a_n} \le y}}} \right\} \to G\left( y \right)$, then $\Pr \left\{ {{{\left( {{{\tilde M}_n} - {b_n}} \right)} \mathord{\left/
 {\vphantom {{\left( {{{\tilde M}_n} - {b_n}} \right)} {{a_n} \le y}}} \right.
 \kern-\nulldelimiterspace} {{a_n} \le y}}} \right\} \to G\left( {ay + b} \right)$ where ${G^\theta }\left( y \right) = G\left( {ay + b} \right)$ for some $a > 0$ and $b$. Consequently, we also have $\Pr \left\{ {{{\left( {{{\tilde M}_n} - {\tilde b_n}} \right)} \mathord{\left/
 {\vphantom {{\left( {{{\tilde M}_n} - {\tilde b_n}} \right)} {{\tilde a_n} \le y}}} \right.
 \kern-\nulldelimiterspace} {{\tilde a_n} \le y}}} \right\} \to G\left( y \right)$, where ${\tilde a_n} = {{{a_n}} \mathord{\left/
 {\vphantom {{{a_n}} a}} \right.
 \kern-\nulldelimiterspace} a}$, and ${\tilde b_n} = {b_n} - {{b{a_n}} \mathord{\left/ {\vphantom {{b{a_n}} a}} \right. \kern-\nulldelimiterspace} a}$.
\end{lemma}

\begin{remark}
Lemma \ref{lemma1} establishes the weak dependence condition $D\left( {{u_n}} \right)$ for a dependent stationary sequence. Lemma \ref{lemma2} highlights that when the $D\left( {{u_n}} \right)$ condition is satisfied, the maximum of a stationary sequence follows an EVD. According to Lemma \ref{lemma3}, the EVD of a dependent stationary sequence belongs to the same type as that of the corresponding i.i.d.~RV sequence. By appropriately modifying the normalizing parameters derived for the i.i.d.~RV sequence, the dependent sequence can be shown to converge to the identical EVD.
\end{remark}

The above conclusions provide the condition under which a stationary sequence converges to the EVD, and clarify the relationship between the EVD of a stationary sequence and that of the corresponding i.i.d.~RV sequence.

To better understand the rationale behind fitting the FAS channel distribution using the EVD, we attempt to analyze the characteristics of the RV sequence $\left\{ {\left| {{h_{\rm{1}}}} \right|, \dots ,\left| {{h_N}} \right|} \right\}$.
\begin{enumerate}
\item It is straightforward to verify that for any $n$ and $m$, the distributions of $\left\{ {\left| {{h_{{j_1}}}} \right|, \dots ,\left| {{h_{{j_n}}}} \right|} \right\}$ and $\left\{ {\left| {{h_{{j_1} + m}}} \right|, \dots ,\left| {{h_{{j_n} + m}}} \right|} \right\}$ are identical. Consequently, the sequence $\left\{ {\left| {{h_{\rm{1}}}} \right|, \dots ,\left| {{h_N}} \right|} \right\}$ is stationary.
\item There exist many pairs ($W$, $N$) for which the $D\left( {{u_n}} \right)$ condition is satisfied when $N$ and $W$ are large\footnote{Under Jakes' model, the channel envelope $X_n=|h_n|$ can be written as
$X_n=\sigma\sqrt{\xi_n^2+\eta_n^2}$, where $\{\xi_n\}$ and $\{\eta_n\}$ are independent stationary
Gaussian sequences with correlation function $r(k)=J_0(2\pi\rho k)$, and $\rho=W/(N-1)$ denotes the physical spacing between any two adjacent ports (in wavelengths).
For any $\rho_0>0$, a standard large-argument bound for the Bessel function (based on Hankel's asymptotic expansion of $J_\nu(x)$ \cite{DLMF})
implies that $|r(k)| \le C_1 k^{-1/2}$ for all $\rho\ge\rho_0$ and all sufficiently large $k$.
Let $I$ and $J$ be two index sets separated by at least $\ell$.
By a Gaussian comparison inequality of the Berman/Li--Shao type, the dependence between
$A_I(u)=\cap_{i\in I}\{X_i\le u\}$ and $A_J(u)=\cap_{j\in J}\{X_j\le u\}$ satisfies
$
\big|\mathbb P(A_I(u)\cap A_J(u))-\mathbb P(A_I(u))\mathbb P(A_J(u))\big|
\le
K\sum_{i\in I}\sum_{j\in J} |r(j-i)|\,e^{-c u^2}.
$
Choose $\ell_N=\lceil N^\beta\rceil$ with $\beta\in(0,1)$ (so that $\ell_N=o(N)$).
Since $\mathbb P(X_1>u_N)=O(1/N)$ and $X_1$ follows a Rayleigh distribution, it follows that $u_N^2=2\sigma^2\log N + O(1)$, and hence
$e^{-c u_N^2}=O(N^{-\delta})$ for some $\delta>0$.
Therefore, the above bound vanishes as $N\to\infty$, yielding the condition $D(u_N)$.
Finally, for any fixed $\rho>0$, choosing $W=\rho(N-1)$ (so that $\rho\ge\rho_0$ for large $N$) ensures that $D(u_N)$ holds for all sufficiently large $N$,
and hence there exist many pairs $(N,W)$ that satisfy the condition.}.
\item Let $\{ |\hat{h}_1|,\dots,|\hat{h}_N|\}$ denote the i.i.d.~RV sequence associated with $\left\{ {\left| {{h_{\rm{1}}}} \right|, \dots ,\left| {{h_N}} \right|} \right\}$, and $| {{{\hat h}_{{\rm{FAS}}}}}| = \max \{ |\hat{h}_1|,\dots,|\hat{h}_N|\}$.
    Given the distribution of $| {{{\hat h}_{i}}} |$, the specific limiting distribution and normalizing parameters can be derived for $| {{{\hat h}_{{\rm{FAS}}}}}|$, which will be shown in Section \ref{sec:gumbel}.
\end{enumerate}
Based on these properties, when the weak dependence condition $D\left( {{u_n}} \right)$ is satisfied and $N \to \infty $, it is reasonable to approximate the distribution of $| {{h_{{\rm{FAS}}}}} |$ using the same limiting distribution of $| {{{\hat h}_{{\rm{FAS}}}}} |$ with the modified normalizing parameters according to Lemma \ref{lemma3}.\footnote{Although the normalizing parameters of $| {{{\hat h}_{{\rm{FAS}}}}}|$ can be derived with given distribution of $| {{{\hat h}_{i}}} |$, the normalizing parameters of $| {{h_{{\rm{FAS}}}}} |$ cannot be obtained directly from the relationships ${\tilde a_n} = {{{a_n}} \mathord{\left/
 {\vphantom {{{a_n}} a}} \right.
 \kern-\nulldelimiterspace} a}$ and ${\tilde b_n} = {b_n} - {{b{a_n}} \mathord{\left/ {\vphantom {{b{a_n}} a}} \right. \kern-\nulldelimiterspace} a}$ in Lemma \ref{lemma3}, since $a$ and $b$ are unknown. Therefore, we fit the normalizing parameters of $| {{h_{{\rm{FAS}}}}} |$ (referred to as the modified normalizing parameters) directly using distribution fitting technique as the identical EVD of $| {{{\hat h}_{{\rm{FAS}}}}} |$. }

It should be emphasized that EVT is used here as a finite-dimensional statistical modeling tool rather than a strict asymptotic result, and the $D(u_N)$ condition is introduced to provide theoretical motivation for employing the EVD to model the FAS channel.
In practical FAS, the number of RVs $N$ is finite, and the $D(u_N)$ condition may not hold for some pairs of $W$ and $N$. However, regardless of whether or not the $D(u_N)$ condition strictly holds, we model the distribution of $| {{h_{{\rm{FAS}}}}}|$ using the identical EVD of $| {{{\hat h}_{{\rm{FAS}}}}} |$ suggested by asymptotic analysis and fit the corresponding normalizing parameters through the distribution fitting technique, rather than adopting the asymptotic normalizing constants derived from EVT. As a result, the effects of spatial correlation and finite $N$ are inherently absorbed into the fitted parameters.
Accordingly, whether the $D(u_N)$ condition holds does not affect the fitted results or the distribution fitting procedure. The primary focus of this paper is to evaluate whether the fitted distribution accurately captures the statistical behavior of the FAS channel, which will be validated through simulations in the following sections.
It is anticipated that the fitted distribution may deviate slightly from the exact channel distribution, while it still provides a reasonable approximation of the FAS channel distribution.
 %

\subsection{Realistic Data Modeling}\label{ssec:data}
To better elucidate the appropriateness of employing the EVD to fit the distribution of ${\left| {{h_{{\rm{FAS}}}}} \right|}$, we analyze this problem from a data-fitting perspective. Although computationally intensive, Monte-Carlo simulations remain the most accurate approach for assessing the FAS performance, since it derives the channel distribution by performing probabilistic statistics on generated Monte-Carlo simulation samples. Here, we arrange the Monte-Carlo simulation samples into an observation sequence, and analyze its characteristics and distribution within the EVT framework.

\begin{figure}
\centering
\includegraphics[width=0.9\columnwidth]{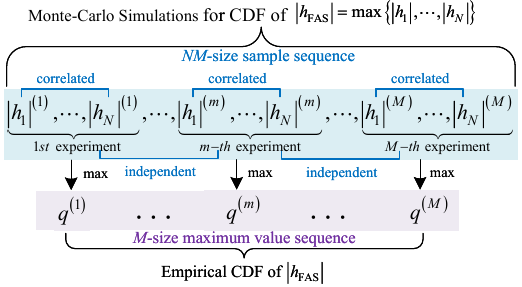}\\
\caption{The flowchart of the Monte-Carlo simulation process.}\label{fig:flowchart}
\end{figure}

\emph{Monte-Carlo Simulation Process for Obtaining the CDF of $\left| {{h_{{\rm{FAS}}}}} \right|$:}
By definition, $\left| {{h_{{\rm{FAS}}}}} \right| = \max \left\{ {\left| {{h_{\rm{1}}}} \right|, \dots ,\left| {{h_N}} \right|} \right\}$, where $\left| {{h_{i}}} \right|$ follows the same distribution, and $\left| {{h_{i}}} \right|$ and $\left| {{h_{j}}} \right|$ are correlated.
To numerically obtain the CDF of $\left| {{h_{{\rm{FAS}}}}} \right|$, we conduct $M$ independent experiments, each involving the generation of $N$ random samples. In the $m$-th experiment, the $N$ samples, denoted by $\{ {{{\left| {{h_{\rm{1}}}} \right|}^{\left( m \right)}}, \dots ,{{\left| {{h_N}} \right|}^{\left( m \right)}}} \}$, are obtained and the maximum value among them, ${q^{\left( m \right)}} = \max \{ {{{\left| {{h_{\rm{1}}}} \right|}^{\left( m \right)}}, \dots ,{{\left| {{h_N}} \right|}^{\left( m \right)}}} \}$, is recorded. Repeating this process over $M$ experiments yields a maximum value sequence, $\{ {{q^{\left( 1 \right)}}, \dots ,{q^{\left( M \right)}}} \}$, from which the empirical CDF of $| h_{\mathrm{FAS}} |$ is estimated, see Fig.~\ref{fig:flowchart}.

The generated $NM$-size samples through the Monte-Carlo simulation process can be viewed as a sample sequence of $NM$ RVs,  divided into $M$ blocks of $N$ variables each, in which variables within each block are correlated and share the same dependence structure, while variables across blocks are independent. On this basis, $NM$ RVs are piecewise stationary sequences, and the maximum value of each block can be extracted and its distribution is subsequently modeled. Such a framework resonates with general methods for modeling the extreme value of the stationary and dependent sequences using the EVD such as time series. Specifically, consider a dependent sequence $\left\{ {{X_1}, \dots ,{X_n}} \right\}$ with observed realizations $\left\{ {{x_1}, \dots ,{x_n}} \right\}$, which is generally divided into $M$ blocks of length $N$. Within each block, there exists a dependence relationship, while maxima of different blocks are nearly independent. Based on the maximum of each block, the EVD is applied to fit the extreme value of the dependent sequences when $N$ is sufficiently large. Accordingly, it is feasible to fit the distribution of $| {{h_{{\rm{FAS}}}}}|$ using the EVD, where the normalizing parameters are fitted from Monte-Carlo simulation samples rather than being computed from the derived formulas for its corresponding i.i.d.~RV sequence.

\begin{remark}
We highlight that there are some discrepancies between the exact distribution of ${\left| {{h_{{\rm{FAS}}}}} \right|}$ and the EVD with the fitted normalizing parameters, due to the following reasons.
\begin{enumerate}
\item The $NM$-RV sequence is piecewise stationary rather than fully stationary. But the RVs within each block share the same distribution and dependence structure, and thus the distribution properties of the block maxima remain consistent with those of a stationary sequence.
\item The dependence in each block generally is weak for modeling, while in FAS, it is determined by the parameter settings. Besides, the length of each block in FAS is a finite, fixed value $N$ rather than sufficiently large.
\end{enumerate}
\end{remark}

\section{Performance Evaluation Using\\the Gumbel Distribution}\label{sec:gumbel}
Here, we first review some results of the classical EVT for i.i.d.~RVs following the Rayleigh distribution. Then, its asymptotic distribution, the Gumbel distribution, is used to fit the distribution of $\left| {{h_{{\rm{FAS}}}}} \right|$ according to the ML criterion. Based on this, the parameters of the Gumbel distribution are fitted as functions of $N$ and $W$. As a consequence, the approximate OP and EC are derived using the fitted Gumbel distribution. Finally, we present numerical results to verify the feasibility of evaluating the FAS performance using the Gumbel distribution.

\subsection{Classical EVT Results}\label{ssec:evtresult}
According to the Fisher-Tippett-Gnedenko theorem, there are only three possible nondegenerate limiting distributions for maxima defined in \eqref{2A}--\eqref{4A}. Now, it is important to figure out which of the three types of limit law applies when each RV ${X_n}$ has a given distribution $F\left( x \right)$, since this determines the exact limiting distribution. Specifically, if a distribution function $F\left( x \right)$ results in one limiting distribution for extremes, it belongs to the attraction domain of this limiting distribution. Next, we introduce the following lemma, providing a sufficient condition for a distribution function belonging to the domain of attraction of the Gumbel distribution \cite{8809213,1710338}.

\begin{lemma}\label{lemma4}
Let $\left\{ {{X_1},{X_2}, \dots ,{X_n}} \right\}$ be an i.i.d.~RV sequence with identical distribution function $F\left( x \right)$. Define $\omega \left( F \right) = \sup \left\{ {x:F\left( x \right) < 1} \right\}$ as the right end-point of the support of $F\left( x \right)$. Assume that there is a real number ${x_1}$ such that for all ${x_1} \le x < \omega \left( F \right)$, $f\left( x \right) = F'\left( x \right) \ne 0$ and $F''\left( x \right)$ exists. If
\begin{align}\label{10A}
\mathop {\lim }\limits_{x \to \omega \left( F \right)} \frac{d}{{dx}}\left[ {\frac{{1 - F\left( x \right)}}{{f\left( x \right)}}} \right] = 0,
\end{align}
then there exist parameters $\left\{ {{a_n} > 0} \right\}$ and $\left\{ {{b_n}} \right\}$ such that ${{\left( {{M_n} - {b_n}} \right)} \mathord{\left/
 {\vphantom {{\left( {{M_n} - {b_n}} \right)} {{a_n}}}} \right. \kern-\nulldelimiterspace} {{a_n}}}$ uniformly converges in distribution to a normalized Gumbel RV as $n \to \infty $. Accordingly, we have $\mathop {\lim }\limits_{n \to \infty } \Pr \left( {{M_n} \le x} \right) = \exp \left( { - \exp \left( { - \frac{{x - {b_n}}}{{{a_n}}}} \right)} \right)$. The normalizing parameters ${a_n}$ and ${b_n}$ (or the scale parameter and location parameter of the Gumbel distribution) are determined by
\begin{equation}\label{11A}
\left\{\begin{aligned}
{a_n}&={F^{ - 1}}\left({1 - \frac{1}{{ne}}} \right)- {F^{ - 1}}\left( {1 -\frac{1}{n}} \right),\\
{b_n}&={F^{ - 1}}\left({1 - \frac{1}{n}} \right),
\end{aligned}\right.
\end{equation}
where ${F^{ - 1}}\left( x \right) = \inf \left\{ {y:F\left( y \right) \ge x} \right\}$.

If $| {{{\hat h}_n}} |$ follows a Rayleigh distribution with CDF ${F_{| {{{\hat h}_n}}|}}\!\left( x \right) \!= \!1\! -\! \exp \left( \!{ - {{{x^2}} \mathord{\left/
 {\vphantom {{{x^2}} {2{\sigma ^2}}}} \right. \kern-\nulldelimiterspace} {2{\sigma ^2}}}}\! \right)$, \eqref{10A} holds for $| {{{\hat h}_n}}|$. According to Lemma \ref{lemma3}, we have
\begin{align}\label{111A}
\mathop {\lim }\limits_{n \to \infty } \Pr \left( |\hat{h}_{\rm FAS}| \le x\right) = \exp \left( { - \exp \left( { - \frac{{x - {{\hat b}_n}}}{{{{\hat a}_n}}}} \right)} \right),
\end{align}
where the normalizing parameters ${\hat a_N}$ and ${\hat b_N}$ (or the scale parameter and location parameter of the GEV distribution) are, respectively, given by
\begin{equation}\label{112A}
\left\{\begin{aligned}
{\hat a_N} &= \frac{\sigma}{\sqrt {2\ln N}},\\
{\hat b_N} &= \sigma \sqrt {2\ln N}.
\end{aligned}\right.
\end{equation}
\end{lemma}

\subsection{Distribution Fitting of $\left| {{h_{{\rm{FAS}}}}} \right|$}\label{ssec:fit-gumbel}
Combining the results in Sections \ref{sec:evd} and \ref{ssec:evtresult}, the distribution of $| {{h_{{\rm{FAS}}}}} |$ can be fitted by the Gumbel distribution under Rayleigh fading, with the modified normalizing parameters rather than the computed normalizing parameters ${\hat a_N}$ and ${\hat b_N}$ in \eqref{112A} for $| {{{\hat h}_{{\rm{FAS}}}}}|$. To this end, we fit the distribution of $| {{h_{{\rm{FAS}}}}}|$ using the Gumbel distribution based on the ML criterion \cite{cousineau2004fitting}. In this process, we can obtain the modified normalizing parameters ${a_N}$ and ${b_N}$, which also correspond to the scale and location parameters of the Gumbel distribution.

Specifically, we first construct the log-likelihood function
\begin{align}\label{80A}
\mathcal{L}\left( {{a_N},{b_N}} \right) = \sum\limits_{p = 1}^{{N_{{\rm{sam}}}}} {\log f\left( {{x_p};{a_N},{b_N}} \right)},
\end{align}
where $f\!\left(\! {x;{a_N},{b_N}}\! \right) \!=\! \frac{1}{{{a_N}}}\!\exp \!\left( \!{ \!- \frac{{x - {b_N}}}{{{a_N}}}}\! \right)\!\exp \!\left[ \!{ - \exp \!\left( \!{ - \frac{{x - {b_N}}}{{{a_N}}}}\! \right)}\! \right]$ is the probability density function (PDF) of the Gumbel distribution, ${{x_p}}$, $p \in \left\{ {1, \dots ,{N_{{\rm{sam}}}}} \right\}$, is a sample of $\left| {{h_{{\rm{FAS}}}}} \right|$, and $N_{\rm{sam}}$ denotes the number of samples. Next, the ML estimation equations can be obtained by taking the partial derivatives of \eqref{80A} with respect to ${a_N}$ and ${b_N}$, respectively, i.e.,
\begin{align}
\frac{{\partial \mathcal{L}}}{{\partial {b_N}}} =& \frac{1}{{{a_N}}}\sum\limits_{p = 1}^{{N_{{\rm{sam}}}}} {\left[ {1 - \exp \left( { - \frac{{{x_p} - {b_N}}}{{{a_N}}}} \right)} \right]}  = 0,\label{81A}\\
\frac{{\partial \mathcal{L}}}{{\partial {a_N}}} =&  - \frac{{{N_{{\rm{sam}}}}}}{{{a_N}}} + \frac{1}{{{{\left( {{a_N}} \right)}^2}}}\sum\limits_{p = 1}^{{N_{{\rm{sam}}}}} {\left( {{x_p} - {b_N}} \right)} \notag\\
 &- \frac{1}{{{{\left( {{a_N}} \right)}^2}}}\sum\limits_{p = 1}^{{N_{{\rm{sam}}}}} {\left( {{x_p} - {b_N}} \right)\exp \left( { - \frac{{{x_p} - {b_N}}}{{{a_N}}}} \right) = 0}.\label{82A}
\end{align}

From \eqref{81A}, the explicit solution of ${{b_N}}$ can be derived as
\begin{align}\label{83A}
{b_N} = -{a_N}\log \left( {\frac{1}{{{N_{{\rm{sam}}}}}}\sum\limits_{p = 1}^{{N_{{\rm{sam}}}}} {{e^{ - {{{x_p}} \mathord{\left/{\vphantom {{{x_p}} {{a_N}}}} \right.
 \kern-\nulldelimiterspace} {{a_N}}}}}} } \right),
\end{align}
while $a_N$ cannot be expressed in closed form according to the above equations. To obtain $a_N$, we substitute \eqref{83A} into \eqref{82A}, yielding the  nonlinear likelihood equation (\ref{84A}) (see top of next page).
\begin{figure*}
\begin{equation}\label{84A}
 - \frac{{{N_{{\rm{sam}}}}}}{{{a_N}}} + \frac{1}{{{{\left( {{a_N}} \right)}^2}}}\sum\limits_{p = 1}^{{N_{{\rm{sam}}}}} {\left( {{x_p} + {a_N}\log \left( {\frac{1}{{{N_{{\rm{sam}}}}}}\sum\limits_{p = 1}^{{N_{{\rm{sam}}}}} {{e^{ - {{{x_p}} \mathord{\left/
 {\vphantom {{{x_p}} {{a_N}}}} \right.
 \kern-\nulldelimiterspace} {{a_N}}}}}} } \right)} \right)}
\left( {1 - \exp \left( { - \left( {\frac{{{x_p}}}{{{a_N}}} + \log \left( {\frac{1}{{{N_{{\rm{sam}}}}}}\sum\limits_{p = 1}^{{N_{{\rm{sam}}}}} {{e^{ - {{{x_p}} \mathord{\left/
 {\vphantom {{{x_p}} {{a_N}}}} \right.
 \kern-\nulldelimiterspace} {{a_N}}}}}} } \right)} \right)} \right)} \right) = 0
\end{equation}
\hrulefill
\end{figure*}
Since it is hard to derive the optimal solution of $a_N$ from \eqref{84A} directly, we solve it using the Newton-Raphson method or another root-finding algorithm to find the optimal numerical solution of $a_N$. Once the optimal $a_N$ is obtained, the optimal $b_N$ can be directly computed from \eqref{83A}.

The process to obtain the distribution of $\left| {{h_{{\rm{FAS}}}}} \right|$ can be summarized as follows:
\begin{itemize}
\item Set the parameters of FAS;
\item Generate the Monte-Carlo simulation samples of $\left| {{h_{{\rm{FAS}}}}} \right|$\footnote{
The channel samples used for distribution fitting are not measured from practical systems; instead, they are generated according to existing channel models $\left| {{h_{{\rm{FAS}}}}} \right|$.
Therefore, practical measurement noise and inaccuracies are beyond the scope of this work.} according to the fully correlated Rayleigh fading channel model described in Section \ref{ssec:model};
\item Fit the scale and location parameters of the Gumbel distribution (corresponding to the normalizing constants ${a_N}$ and ${b_N}$) according to the ML criterion, which can be done by using the MATLAB function {\tt evfit};
\item Finally, compute the CDF of $\left| {{h_{{\rm{FAS}}}}} \right|$ using
$${F_{\left| {{h_{{\rm{FAS}}}}} \right|}}\left( x \right) = \exp \left( { - \exp \left( { - \frac{{x - {b_N}}}{{{a_N}}}} \right)} \right).$$
\end{itemize}
Following the above process, the Gumbel distribution parameters can be obtained for given $W$ and $N$, while the fitting process must be carried out again when $W$ and $N$ change. Moreover, these parameters are fitted based on the Monte-Carlo simulation samples of $| {{h_{{\rm{FAS}}}}}|$. As $N$ increases, the computational complexity grows. Therefore, this fitting approach may not be suitable for practical engineering applications.

\subsection{Parameter Estimation}\label{ssec:fit-gumbel-parameter}

To facilitate efficient performance evaluation, we introduce an explicit mapping between the system parameters $(W,N)$ and the distribution parameters. The objective of this mapping is to obtain simple yet sufficiently accurate analytical expressions for computing the distribution parameters.
Specifically,
the procedure of fitting these parameters as functions of $W$ and $N$ is summarized as follows.
\begin{itemize}
  \item Under a wide range of system parameters $W$ and $N$, Monte-Carlo simulations are performed to generate samples of the FAS channel $\left|h_{\rm{FAS}}\right|$ for each pair ($W$, $N$).
  \item For each pair ($W$, $N$), the parameters of the Gumbel and GEV distributions are estimated using the MATLAB function {\tt evfit} according to the ML criterion.
  \item The estimated distribution parameters obtained for all ($W$, $N$) combinations are then fitted using multivariate polynomial regression with respect to $W$ and $N$.
\end{itemize}
By doing so, we obtain
\begin{align}
{a_N} \approx &~ 3.928\times {10^{ - 1}} - 3.528\times {10^{ - 2}}W + 9.585\times{10^{ - 4}}N\notag\\
 & +2.817\times{10^{ -3}}{W^2}+ 3.703\times{10^{ -4}}WN\notag\\
 & - 2.94\times {10^{ - 5}}{N^2}- 4.659\times{10^{ - 5}}{W^2}N\notag\\
 & + 8.07 \times {10^{ - 7}}W{N^2}+ 1.289 \times {10^{ - 7}}{N^3},\label{21A}\\
{b_N} \approx &~ 9.261\times {10^{ - 1}} + 2.629\times {10^{ - 1}}W + 7.106\times {10^{ - 3}}N\notag\\
 & - 3.35\times {10^{ - 2}}{W^2}-8.59\times {10^{ -4}}WN\notag\\
 & - 9.37\times {10^{ - 5}}{N^2}+ 4.863\times {10^{ -4}}{W^2} N\notag\\
 & - 2.84\times {10^{ - 5}}W{N^2}+1.192 \times {10^{ - 6}}{N^3}.\label{22A}
\end{align}
From \eqref{21A} and \eqref{22A}, ${a_N}$ and ${b_N}$ can be quickly computed under varying $W$ and $N$, which significantly reduces the computational complexity of parameter estimation and consequently, the evaluation of OP and EC based on the Gumbel distribution. The above parameter fitting is carried out within the ranges $W \in \left[ {0.5,5} \right]$ and $\rho  \in \left[ {0.05,0.5} \right]$, where $\rho  = {W \mathord{\left/
 {\vphantom {W {\left( {N - 1} \right),}}} \right.
 \kern-\nulldelimiterspace} {\left( {N - 1} \right)}}$ denotes the physical spacing between any two adjacent ports in units of the wavelength with $N > 1$. The considered range of $W$ largely encompasses the antenna sizes commonly adopted in existing representative studies \cite{10103838, 10678877, 9264694, wong2022closed,10623405}.
Since a half-wavelength spacing is sufficient to ensure weakly correlated channels, $0.5$ is chosen as the upper bound of $\rho$.
Moreover, the existing research \cite{10103838, wong2022closed,10623405} shows that increasing $N$ under a fixed $W$
yields diminishing performance gains of FASs.
Besides, increasing $N$ beyond a certain point not only provides marginal additional gains, but also substantially increases the channel estimation overhead and hardware complexity.
Therefore, we choose a relatively small value of 0.05 as the lower bound of $\rho$, which is sufficiently large for
the system performance to approach saturation.
This range is broad enough to cover both dense port deployment with compact antenna sizes and sparse port deployment with large antenna sizes.

Polynomial parameterization is adopted as a representation of this mapping due to the following reasons.
Mathematically, any continuous mapping on a compact set (i.e., a closed and bounded set in $\mathbb{R}^d$) can be uniformly approximated arbitrarily well by polynomials, by the Stone-Weierstrass theorem \cite{Folland1999}. However, our intent is not to achieve arbitrarily high fitting accuracy through high-order expansions. Instead, we deliberately select a low-order polynomial to balance approximation accuracy, numerical robustness, and analytical tractability.
While it is possible to impose a fixed functional form based solely on physical intuition, enforcing such a rigid structure may introduce numerical sensitivity and reduced accuracy over a finite parameter range.

During parameter estimation, to obtain a unified and analytically tractable channel representation, we focus on modeling normalized small-scale fading, and derive the corresponding distribution parameters in \eqref{21A} and \eqref{22A} without explicitly considering large-scale fading or different fading severities. Nevertheless, the resulting normalized small-scale channel distribution can still be used to describe channels in different environments. Specifically, for a transmitter-receiver separation distance $d$ and a path-loss exponent $\alpha$, large-scale fading can be modeled as $d^{-\alpha}$. For small-scale fading, we assume the channel coefficients follow $\mathcal{CN}(0,\sigma^2)$, where $\sigma^2$ denotes the channel power. As such, a practical FAS channel distribution can be computed by
$\Pr \left( {\sqrt {{d^{ - \alpha }}{\sigma ^2}} \left| {{h_{{\rm{FAS}}}}} \right| \le r} \right) = \Pr \left( {\left| {{h_{{\rm{FAS}}}}} \right| \le \frac{r}{{\sqrt {{d^{ - \alpha }}{\sigma ^2}} }}} \right)$ using the fitted channel distribution.

\subsection{Approximate OP and EC under Rayleigh Fading}
Based on the above, we provide the following theorems to compute the OP and EC of FAS under Rayleigh fading.

\begin{theorem}\label{theorem:op}
The approximate OP for FAS under Rayleigh fading using the Gumbel distribution is given by
\begin{align}\label{40A}
 {P_{{\rm{out}}}} = \exp \left( { - \exp \left( { - \frac{{\hat \gamma  - {b_N}}}{{{a_N}}}} \right)} \right).
\end{align}
\end{theorem}

\emph{Proof:} Since the CDF of ${\left| {{h_{{\rm{FAS}}}}} \right|}$ can be approximately fitted as ${F_{\left| {{h_{{\rm{FAS}}}}} \right|}}\left( x \right) = \exp \left( { - \exp \left( { - \frac{{x - {b_N}}}{{{a_N}}}} \right)} \right)$, substituting the SNR threshold ${\hat \gamma }$ from \eqref{8A} into ${F_{\left| {{h_{{\rm{FAS}}}}} \right|}}\left( x \right)$ yields \eqref{40A}.\hfill {$\blacksquare $}

\begin{theorem}\label{theorem:ec}
The approximate EC for FAS under Rayleigh fading using the Gumbel distribution is given by
\begin{align}\label{41A}
\bar C = {{e}_N}{\gamma} + {{d}_N}\;\;\;{{{{{\rm{nats}}} \mathord{\left/
 {\vphantom {{{\rm{nats}}} {\rm{s}}}} \right.
 \kern-\nulldelimiterspace} {\rm{s}}}} \mathord{\left/
 {\vphantom {{{{{\rm{nats}}} \mathord{\left/
 {\vphantom {{{\rm{nats}}} {\rm{s}}}} \right.
 \kern-\nulldelimiterspace} {\rm{s}}}} {{\rm{Hz}}}}} \right.
 \kern-\nulldelimiterspace} {{\rm{Hz}}}},
\end{align}
where ${\gamma} = 0.5772 \cdots $ is the Euler constant. In \eqref{41A}, ${{e}_N}$ and ${{d}_N}$ are given, respectively, by
 \begin{align}
{{e}_N} &= {\ln}\left( {1 + {\alpha _N}\bar \gamma  + {\beta _N}\bar \gamma } \right) - {\ln}\left( {1 + {\alpha _N}\bar \gamma } \right),\label{44A}\\
{{d}_N} &= {\ln}\left( {1 + {\beta _N}\bar \gamma } \right),\label{45A}
\end{align}
where ${\alpha _N} = 2{{a}_N}{{ b}_N}$ and ${\beta _N} = {{ b}_N}{{ b}_N}$.
\end{theorem}

\emph{Proof:} According to the limiting throughput theorem in \cite{1710338}, ${{\left( {\hat C - {{\hat d}_N}} \right)} \mathord{\left/
 {\vphantom {{\left( {\hat C - {{\hat d}_N}} \right)} {{{\hat e}_N}}}} \right. \kern-\nulldelimiterspace} {{{\hat e}_N}}}$ converges uniformly in distribution to a normalized Gumbel RV, where $\hat C = {\ln ( {1 + {{| {{{\hat h}_{{\rm{FAS}}}}} |}^2}\bar \gamma } )}$. The normalizing parameters ${{{\hat e}_N}}$ and ${{{\hat d}_N}}$ are, respectively, given by
 \begin{align}
{{\hat e}_N}& = {\ln}\left( {1 + {\hat \alpha _N}\bar \gamma  + {\hat \beta _N}\bar \gamma } \right) - {\ln}\left( {1 + {\hat \alpha _N}\bar \gamma } \right),\label{42A}\\
{{\hat d}_N} &= {\ln}\left( {1 + {\hat \beta _N}\bar \gamma } \right),\label{43A}
\end{align}
where ${\hat \alpha _N} = 2{{\hat a}_N}{{\hat b}_N}$ and ${\hat \beta _N} = {{\hat b}_N}{{\hat b}_N}$ denotes the normalizing parameters of ${{{| {{{\hat h}_{{\rm{FAS}}}}} |}^2}}$, which also converges to the Gumbel distribution.\footnote{Under Rayleigh fading, $|\hat{h}_n|^2$ follows the exponential distribution with parameter $\lambda$. It is easy to check that the CDF of $|\hat{h}_n|^2$, denoted by $F_{|h_n|^2}(x)$, satisfies \eqref{10A}. Therefore, $(|\hat{h}_{\rm FAS}|^2 - \hat{\beta}_N)/\hat{\alpha} _N$ converges uniformly in distribution to a normalized Gumbel RV, where $\hat{\alpha}_N =1/\lambda$ and $\hat{\beta}_N= \ln  N /\lambda$. According to ${\hat a_N}$, ${\hat b_N}$, ${{\hat \alpha _N}}$ and ${{\hat \beta _N}}$, we have ${\hat \alpha _N} = 2{{\hat a}_N}{{\hat b}_N}$ and ${\hat \beta _N} = {{\hat b}_N}{{\hat b}_N}$, which reflect the relationship between the normalizing parameters of $|\hat{h}_{\rm FAS}|$ and that of $|\hat{h}_{\rm FAS}|^2$.} Accordingly, we have $(\mathbb{E}\{\hat{C}\} - \hat{d}_N)/ \hat{e}_N \to \gamma$ as $N \to \infty$. For a large $N$, $\mathbb{E}\{\hat C\}$ can be approximated by using $\mathbb{E}\{\hat C\} \approx {\hat e_N}{\gamma} + {\hat d_N}$.

According to Section \ref{sec:evd}, the EC of FAS ${\bar C}=\mathbb{E}\left\{ {C} \right\}$, in which $C = {\ln ( {1 + {{\left| {{{ h}_{{\rm{FAS}}}}} \right|}^2}\bar \gamma })}$. In Section \ref{ssec:data}, we fit the distribution of $\left|h_{\rm{FAS}}\right|$ using the Gumbel distribution with the normalizing parameters ${a_N}$ and ${b_N}$. According to the parameter relationships derived for the i.i.d. sequences above, the distributions of ${{\left| {{{ h}_{{\rm{FAS}}}}} \right|}^2}$ and $C$ can be approximately fitted by the Gumbel distribution with the normalizing parameters ${\alpha _N}$ and ${\beta _N}$, ${e _N}$ and ${d _N}$, respectively. On the other hand, the parameter relationships can be approximated as \eqref{44A} and \eqref{45A}.
Based on the above, we have ${{\left( {\mathbb{E}\left\{C \right\} - {d_N}} \right)} \mathord{\left/ {\vphantom {{\left( {\mathbb{E}\left\{ C \right\} - {d_N}} \right)} {{e_N}}}} \right. \kern-\nulldelimiterspace} {{e_N}}} \to {\gamma}$ as $N \to \infty$. For a large $N$, $\mathbb{E}\left\{C \right\}$ can be approximated by $\bar{C}=\mathbb{E}\left\{C \right\} \approx {e_N}{\gamma} + {d_N}$.\hfill {$\blacksquare $}

In conclusion, our proposed performance evaluation framework consists of an offline parameter-fitting stage and an online performance-computation stage.
In the offline stage, the distribution parameters are estimated from Monte-Carlo generated channel samples via distribution fitting, as described in Section \ref{ssec:fit-gumbel}. Then, we fit the distribution parameters as functions of the system parameters as described in Section \ref{ssec:fit-gumbel-parameter}, which incurs a one-time cost for the considered deployment settings.
In the online stage, once the fitted parameter expressions are obtained, the OP and EC can be evaluated by substituting them into the closed-form expressions in \eqref{40A} and \eqref{41A}, resulting in a very low computational complexity that is independent of the number of ports $N$.

\begin{figure*}[htbp]
\centering
\begin{minipage}[t]{0.64\linewidth}
\centering
\includegraphics[width=0.48\linewidth]{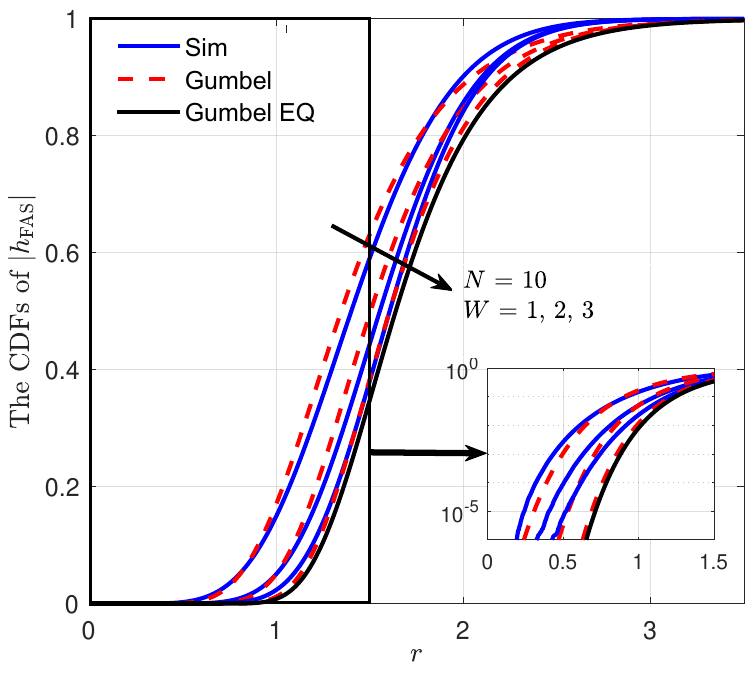}
\hfill
\includegraphics[width=0.48\linewidth]{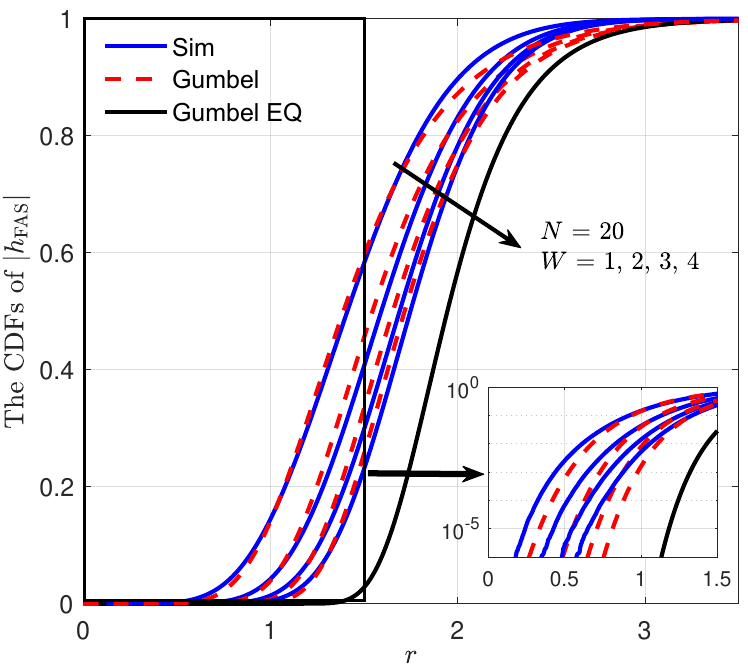}
\caption{The CDFs of $|h_{\rm FAS}|$ obtained from the Monte-Carlo simulations and the fitted Gumbel distribution under $N=10$ and $N=20$, respectively, for selected values of $W$.}\label{fig:cdf-gumbel}
\end{minipage}%
\hfill
\begin{minipage}[t]{0.33\linewidth}
\centering
\includegraphics[width=0.9\linewidth]{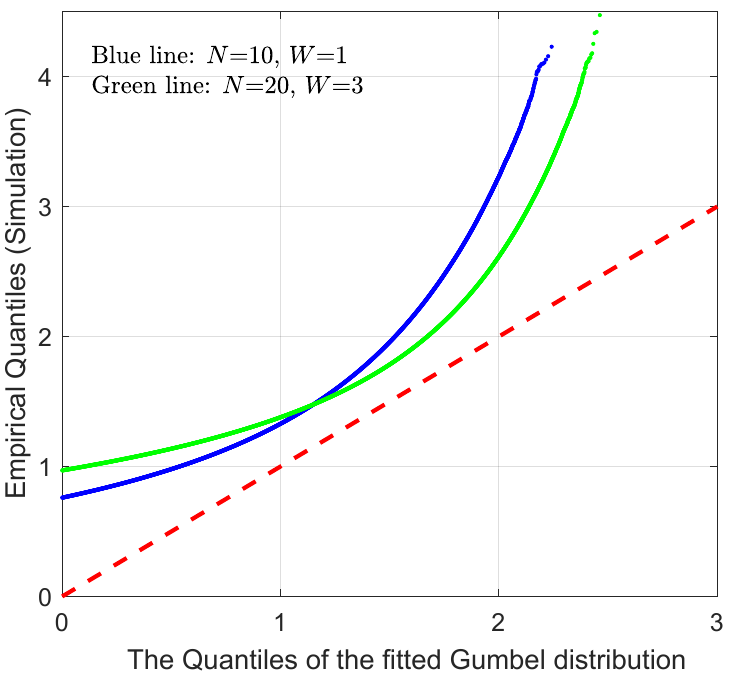}
\caption{The Q-Q plot of the empirical distribution versus the fitted Gumbel distribution.}\label{fig:QQ}
\end{minipage}
\end{figure*}

\subsection{Simulation Results}

\subsubsection*{The accuracy of the fitted Gumbel distribution}
The results in Fig.~\ref{fig:cdf-gumbel} are provided for the CDFs of ${{{\left| {{h_{{\rm{FAS}}}}} \right|}}}$ obtained from 1) the Monte-Carlo simulations\footnote{
It is worth noting that even under ideal channel models, ML estimators can exhibit increased variance in the ultra-low outage regime when the number of available samples is limited.
Hence, we generate a number of channel samples that is several orders of magnitude larger than the inverse of the target OP. To accurately characterize OPs on the order of ${10^{ - 5}}$, we employ up to $10^8$ Monte-Carlo samples to ensure a sufficient number of tail realizations and to reduce estimator variance.} marked as ``Sim", where the channel samples of $\left| {{h_{{\rm{FAS}}}}} \right|$ are generated according to the fully correlated Rayleigh fading channel model described in Section \ref{ssec:model}; 2) the fitted Gumbel distribution with the parameters calculated by \eqref{21A} and \eqref{22A} marked as ``Gumbel", 3) the Gumbel distribution with the parameters computed by \eqref{112A} for i.i.d.~RV sequence, marked as ``Gumbel\_EQ".
In each subfigure, given $N$, the variation of $W$ determines the port deployment density, which reflects the dependence structure among channels.
The black boxes in Fig. 2 indicate that the values in this region are shown in logarithmic scale in the small figures on the right. Overall, the Gumbel distribution provides a good fit of the CDF of ${{{\left| {{h_{{\rm{FAS}}}}} \right|}}}$ across varying $N$ and dependence strengths. However, some deviations are observed in the high-probability and extremely low-probability regions. Since the outage performance generally concentrates on the region below ${10^{ - 2}}$, the deviations in extremely low-probability region may result in noticeable gaps between the approximate OP and Monte-Carlo simulation results. Additionally, the EC, being an average metric, is insensitive to deviations in the low-probability region, but deviations in the high-probability region may introduce small gaps between the approximate EC and the exact one. Moreover, for arbitrary $N$, the CDFs of ${{{\left| {{h_{{\rm{FAS}}}}} \right|}}}$ gradually approach the CDFs of $|\hat{h}_{\rm FAS}|$ as $W$ increases (i.e., as the channel correlation weakens). This indicates that the CDF of ${{{\left| {{h_{{\rm{FAS}}}}} \right|}}}$ can be fitted by the identical Gumbel distribution of the corresponding i.i.d.~RV sequence with parameters ${\hat a_N}$ and ${\hat b_N}$, when the correlation is sufficiently weak satisfying the weak dependence conditions, as discussed in Section \ref{ssec:weak}.

To more intuitively illustrate the fitting error of the Gumbel distribution, Fig.~\ref{fig:QQ} shows the quantile-quantile (Q-Q) plot of the empirical distribution obtained by Monte-Carlo simulation samples of ${{{\left| {{h_{{\rm{FAS}}}}} \right|}}}$ against the fitted Gumbel distribution ${F_{\left| {{h_{{\rm{FAS}}}}} \right|}}\left( x \right)$. The red dashed line is the reference line which represents the empirical quantiles, and the blue and green lines denote the quantiles of the fitted Gumbel distribution under different $N$ and $W$. The Q-Q plot lying close to the red diagonal line indicates a good fit of the Gumbel distribution to the empirical distribution, whereas deviations from the line reflect a poor fit. We observe from Fig.~\ref{fig:QQ} that some deviations exist in the low- and high-quantile regions, which is consistent with the observations in Fig.~\ref{fig:cdf-gumbel}. In simulation, both the number of RVs $N$ and the number of simulation realizations are finite, the minimum of the generated channel samples is inherently limited. Since the parameters of the Gumbel distribution are fitted from the Monte-Carlo simulation samples, the left tail of the fitted Gumbel distribution converges to a finite minimum value, leading to deviations in the low-quantile region. Similarly, due to the finite number of RVs $N$, the distribution of the maximum has not fully converged to the Gumbel distribution, resulting in deviations in the high-quantile region.

\subsubsection*{The accuracy of the approximate OP and EC}
\begin{figure}
\centering
\includegraphics[width=0.9\columnwidth]{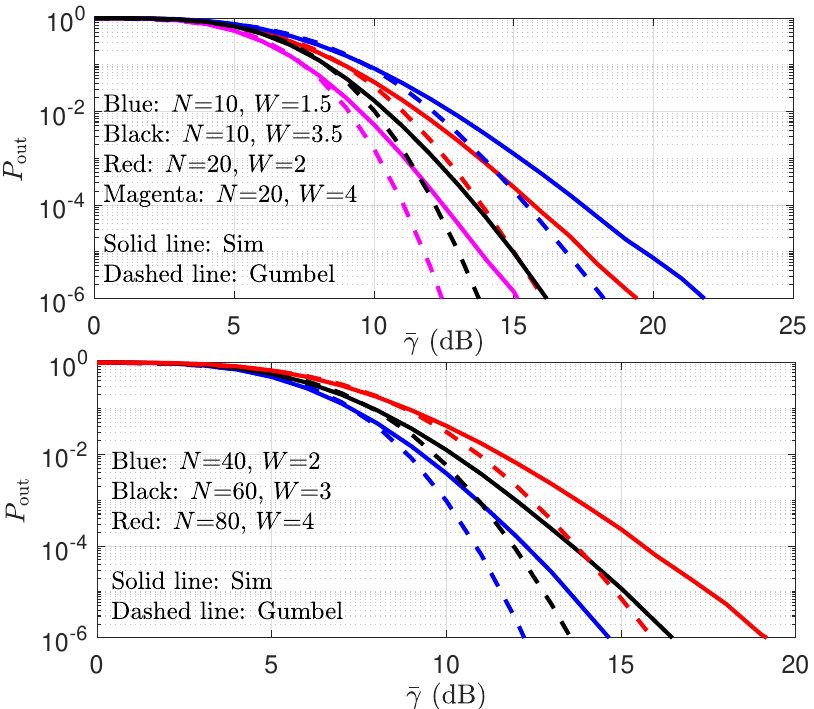}\\
\caption{The OPs of FAS obtained from Monte-Carlo simulations and the fitted Gumbel distribution versus the transmit SNR $\bar \gamma$ for selected values of $W$ and $N$.}\label{fig:op}
\end{figure}

\begin{figure}
\centering
\includegraphics[width=0.9\columnwidth]{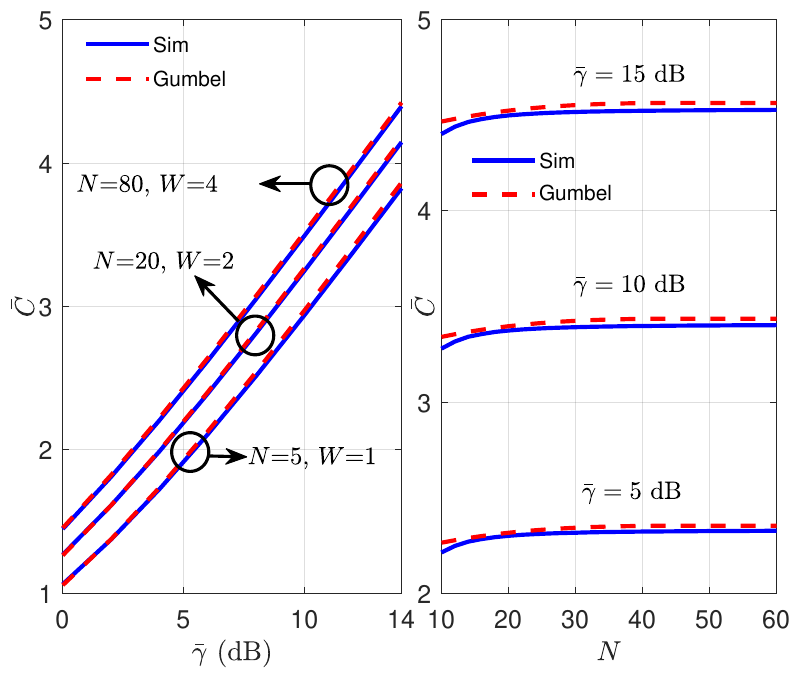}\\
\caption{The ECs of FAS obtained from Monte-Carlo simulations and the fitted Gumbel distribution versus the transmit SNR $\bar \gamma$ and the number of ports $N$, respectively.}\label{fig:ec}
\end{figure}

Now, Fig.~\ref{fig:op} illustrates the approximate OPs of FAS versus the transmit SNR ${\bar \gamma }$ with ${\gamma _{{\rm{th}}}} = 10$ dB. Overall, the approximate OPs of FAS using the fitted Gumbel distribution are accurate in the high-OP region, while some discrepancies occur in the low-OP region. This phenomenon indicates that the fitted Gumbel distribution does not accurately capture the extreme-tail behavior of the FAS channel. Furthermore, Fig.~\ref{fig:ec} shows the approximate ECs of FAS versus the transmit SNR ${\bar \gamma }$ and the number of ports $N$. The approximate ECs using the fitted Gumbel distribution well match the exact ECs under different values of $N$ and $W$. This is because EC, being an average metric, is insensitive to deviations in the extreme lower tail (e.g., at the ultra-low-probability region). Nevertheless, minor deviations between the approximate and exact ECs in the right subfigure arise from the imperfect fit of the Gumbel distribution in the high-probability region. Moreover, in the right subfigure, the ECs approach saturation as $N$ increases, indicating a performance plateau for FAS with a given antenna size $W$. In addition, the ECs increase with higher transmit SNR ${\bar \gamma }$ as expected.

\section{Performance Evaluation Using\\ the GEV Distribution}\label{sec:gev}
From the simulation results, it is known that the Gumbel distribution can provide a reasonable fit for the distribution of $\left| {{h_{{\rm{FAS}}}}} \right|$ in general. However, small deviations exist in the high-probability and ultra-low-probability regions, leading to small gaps between the approximate OP and the simulation results. These deviations are attributed to several factors. First of all, the number of RVs, $N$, is finite, which limits the coverage of extreme values in the Monte-Carlo simulation samples, resulting in small deviations between the actual channel distribution and the fitted Gumbel distribution. Second, there are correlations among the channels, and the amount of correlation is dictated by $N$ and $W$. Different $N$ and $W$ result in different kurtosis and skewness of the actual FAS channel distribution. However, the Gumbel distribution has fixed skewness and kurtosis \cite{gumbel1958statistics}, so it would not fully capture these variations induced by finite sample size and correlations, leading to the observed discrepancies.

To improve the fitting accuracy, in this section, we fit the distribution of ${\left| {{h_{{\rm{FAS}}}}} \right|}$ using a more flexible GEV distribution, which incorporates a shape parameter $\xi$ to adjust the EVD type and has more flexible skewness and kurtosis to control the symmetry and tail behavior. In what follows, we first introduce the process of fitting the distribution of $\left| {{h_{{\rm{FAS}}}}} \right|$ and the parameter estimation for the GEV distribution. Then we evaluate the OP and EC using the fitted GEV distribution. Finally, we present simulation results to show the superior fitting accuracy of the GEV distribution over the Gumbel distribution, and to verify its advantages in terms of approximation accuracy and computational complexity compared to the existing works.

\subsection{Distribution Fitting of $\left| {{h_{{\rm{FAS}}}}} \right|$ and Parameter Estimation}\label{ssec:fit-gev}
Here, the distribution of ${\left| {{h_{{\rm{FAS}}}}} \right|}$ is modeled using the GEV distribution, with the parameters estimated based on the ML criterion, including the shape parameter $\xi$, and the scale and location parameters (corresponding to the modified normalizing parameters $a_N$ and $b_N$) of the GEV distribution.

Similar to Section \ref{ssec:fit-gumbel}, the log-likelihood function can be constructed as
 \begin{align}\label{90A}
\mathcal{L}\left( {\xi ,{a_N},{b_N}} \right) = \sum\limits_{p = 1}^{{N_{{\rm{sam}}}}} {\log f\left( {{x_p};\xi ,{a_N},{b_N}} \right)},
 \end{align}
 where $f\!\left( \!{x;\xi\! ,\!{a_N},\!{b_N}} \right) \!\!= \!\!\frac{1}{{{a_N}}}\!{\left[\! {1\! + \! \xi \frac{{x - {b_N}}}{{{a_N}}}} \!\right]\!^{ \!- 1 \!-\! \frac{1}{\xi }}}\!\!\exp \!\left\{ {\! - \! {{\left[\! {1\! + \!\xi \frac{{x - {b_N}}}{{{a_N}}}} \!\right]}\!^{ - \frac{1}{\xi }}}} \right\}$, for $1 + \xi \frac{{x - {b_N}}}{{{a_N}}} > 0$, denotes the PDF of the GEV distribution. According to \eqref{90A}, we derive the ML estimation equations as ${{\partial \mathcal{L}} \mathord{\left/ {\vphantom {{\partial \mathcal{L}} {\partial {a_N}}}} \right. \kern-\nulldelimiterspace} {\partial {a_N}}} = 0$, ${{\partial \mathcal{L}} \mathord{\left/ {\vphantom {{\partial \mathcal{L}} {\partial {b_n}}}} \right. \kern-\nulldelimiterspace} {\partial {b_n}}} = 0$, and ${{\partial \mathcal{L}} \mathord{\left/ {\vphantom {{\partial \mathcal{L}} {\partial \xi }}} \right. \kern-\nulldelimiterspace} {\partial \xi }} = 0$, which cannot be solved explicitly in closed form. As a result, we resort to solving these equations numerically using nonlinear optimization algorithms to find the parameter estimates.

The fitting process of the distribution of $\left| {{h_{{\rm{FAS}}}}} \right|$ using the GEV distribution is summarized as follows:
\begin{itemize}
\item Set the parameters of FAS;
\item Generate the Monte-Carlo simulation samples of $\left| {{h_{{\rm{FAS}}}}} \right|$;
\item Fit the shape parameter $\xi$, and the scale and location parameters (corresponding to the modified normalizing parameters ${{\tilde a}_N}$ and ${{\tilde b}_N}$) of the GEV distribution based on the ML criterion, using the MATLAB function {\tt gevfit};
\item Compute the CDF of $\left| {{h_{{\rm{FAS}}}}} \right|$ using
\begin{multline*}
{F_{\left| {{h_{{\rm{FAS}}}}} \right|}}\left( x \right) = \exp \left( {  - {{\left( {1 + \xi \frac{{x  - {{\tilde b}_N}}}{{{{\tilde a}_N}}}} \right)}^{ - {1 \mathord{\left/
 {\vphantom {1 \xi }} \right.
 \kern-\nulldelimiterspace} \xi }}}} \right),\\
\mbox{for }1 + \xi \frac{{x - {{\tilde b}_N}}}{{{{\tilde a}_N}}} > 0.
\end{multline*}
\end{itemize}

Similar to the Gumbel distribution fitting, to improve computational efficiency, we estimate the parameters of the GEV distribution as functions of $N$ and $W$ under $W \in \left[ {0.5,5} \right]$ and $\rho  \in \left[ {0.05,0.5} \right]$ based on the ML criterion, and obtain
\begin{align}
\xi  \approx & - 1.235\times{10^{-1}}+ 1.014\times{10^{-3}}W - 8.942\times {10^{ - 6}}N\notag\\
 &+7.796\times{10^{-4}}{W^2} - 8.619\times {10^{ - 5}}WN \notag\\
 &+1.867\times {10^{ - 6}}{N^2}+1.867\times {10^{ - 6}}{W^2}N\notag\\
 & + 2.332\times {10^{ - 6}}W{N^2} -6.288\times {10^{ - 8}}{N^3},\label{23A}\\
{\tilde a_N} \approx &~ 4.039\times {10^{ -1}} - 3.814\times {10^{ -2}}W +8.851\times {10^{ -4}}N\notag\\
 &+3.338\times {10^{ -3}}{W^2}+ 3.779\times {10^{ -4}}WN\notag\\
 &-2.798\times {10^{ - 5}}{N^2} - 5.65\times {10^{ - 5}}{W^2}N\notag\\
 & + 1.552\times {10^{ - 6}}W{N^2} +1.004\times {10^{ - 7}}{N^3},\label{24A}\\
{\tilde b_N} \approx &~ 9.346\times {10^{ - 1}} + 2.511\times {10^{ - 1}}W+9.196\times {10^{ - 3}}N\notag\\
 &- 3.177\times {10^{ - 2}}{W^2} - 6.431\times {10^{ - 4}}WN\notag\\
 & -1.44\times {10^{ - 4}}{N^2}+4.325\times {10^{ - 4}}{W^2}N \notag\\
 & - 2.548 \times {10^{ - 5}}W{N^2}+1.404\times {10^{ - 6}}{N^3}.\label{25A}
\end{align}

In this work, the fitted parameters are obtained under the fully correlated channel model, which provides a theoretically accurate characterization of spatial correlation for FAS.
Although other correlation models have been adopted in the literature, such as the block-diagonal or reference correlation models, these models can be viewed as structured simplifications introduced to facilitate analytical tractability compared to the fully correlated channel model. It should be emphasized that the channel realizations used in our work are generated from a theoretical correlation model rather than measurement-based data.
To the best of our knowledge, there are no publicly available large-scale measurement datasets for FAS that would enable distribution fitting based on real-world channel statistics.
Therefore, for the single-antenna FAS, the fully correlated channel model provides a general and accurate theoretical correlation characterization, and the fitted parameters derived under this model can effectively capture the corresponding spatial correlation.
\subsection{Approximate OP and EC under Rayleigh Fading}
Following Section \ref{ssec:fit-gev}, we provide the following theorems to evaluate the OP and EC under Rayleigh fading of FAS.

\begin{theorem}\label{theorem:op-gev}
The approximate OP for FAS under Rayleigh fading using the GEV distribution fitting is given by
\begin{multline}\label{46A}
{P_{{\rm{out}}}} = \exp \left( {  - {{\left( {1 + \xi \frac{{\hat \gamma  - {{\tilde b}_N}}}{{{{\tilde a}_N}}}} \right)}^{ - {1 \mathord{\left/
 {\vphantom {1 \xi }} \right.
 \kern-\nulldelimiterspace} \xi }}}} \right),\\
\mbox{for }1 + \xi \frac{{\hat \gamma - {{\tilde b}_N}}}{{{{\tilde a}_N}}} > 0.
\end{multline}
\end{theorem}

\emph{Proof:} Since the CDF of ${\left| {{h_{{\rm{FAS}}}}} \right|}$ can be approximately fitted as ${F_{\left| {{h_{{\rm{FAS}}}}} \right|}}\left( x \right)$. By substituting the SNR threshold ${\hat \gamma }$ from \eqref{8A} into ${F_{\left| {{h_{{\rm{FAS}}}}} \right|}}\left( x \right)$, we then obtain \eqref{46A}.\hfill {$\blacksquare $}

\begin{figure*}[htbp]
\centering
\begin{minipage}[t]{0.64\linewidth}
\centering
\includegraphics[width=0.48\linewidth]{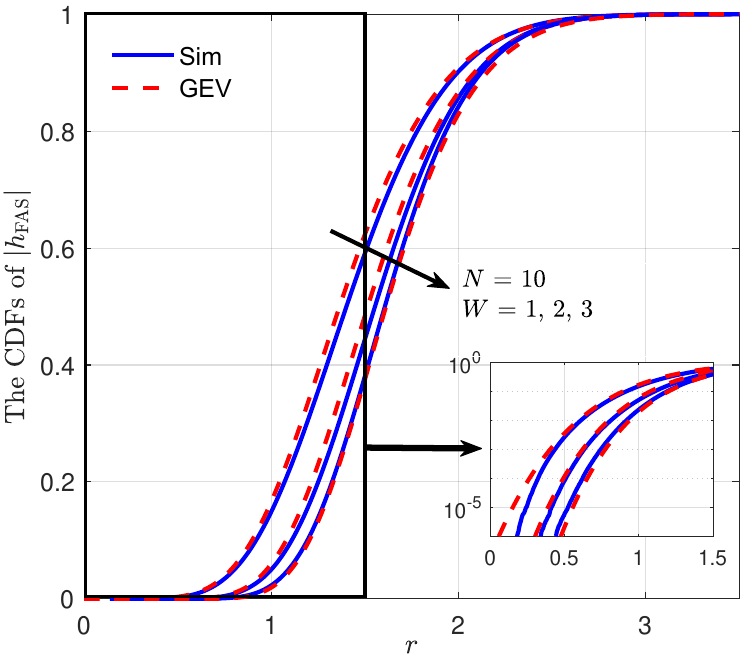}
\hfill
\includegraphics[width=0.48\linewidth]{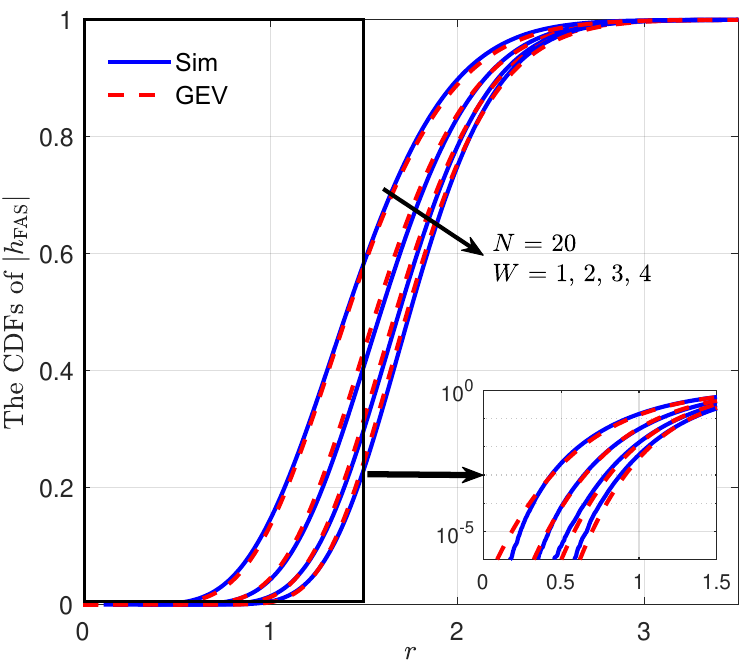}
\caption{The CDFs of $|h_{\rm FAS}|$ obtained from Monte-Carlo simulations and the fitted GEV distribution under $N=10$ and $N=20$, respectively, for selected values of $W$.}\label{fig:cdf-gev}
\end{minipage}%
\hfill
\begin{minipage}[t]{0.33\linewidth}
\centering
\includegraphics[width=0.9\linewidth]{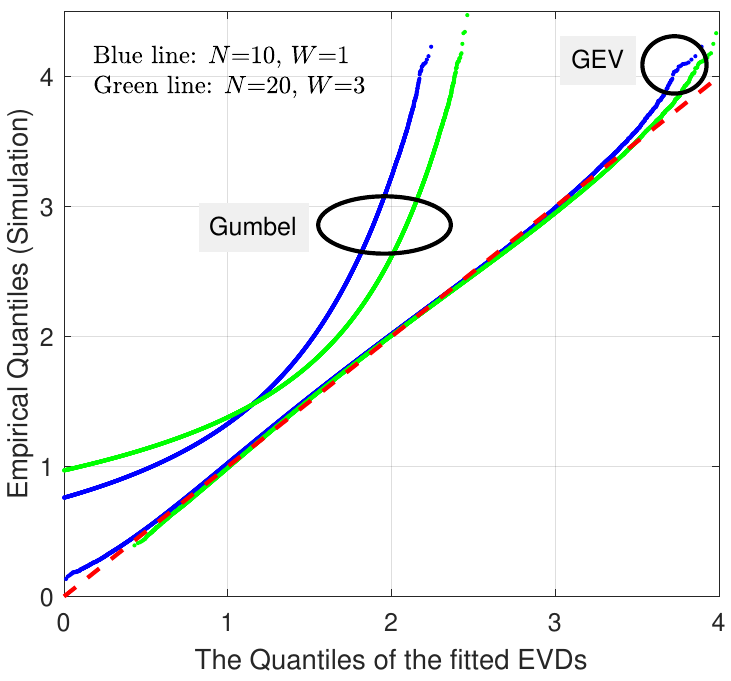}
\caption{The Q-Q plot of the empirical distribution versus the fitted EVDs.}\label{fig:QQ-gev}
\end{minipage}
\end{figure*}

\begin{theorem}\label{theorem:ec-gev}
Based on the mean of a GEV RV, the approximate EC for FAS under Rayleigh fading using the GEV distribution fitting is given by
 \begin{align}\label{47A}
\bar C = \left\{ {\begin{array}{*{20}{c}}
{{{\tilde e}_N} + {{\tilde d}_N}\gamma, \;\;\;\;\;{\rm{if\;}}\tilde \xi  = 0,}\\
{{{\tilde e}_N} + {{{{\tilde d}_N}\left( {{g_1} - 1} \right)} \mathord{\left/
 {\vphantom {{{{\tilde d}_N}\left( {{g_1} - 1} \right)} {\xi ,{\rm{   if\;}}\xi  \ne 0,\tilde \xi  < 1,}}} \right.
 \kern-\nulldelimiterspace} {\xi ,{\rm{   if \;}}\tilde \xi  \ne 0,\tilde \xi  < 1,}}}\\
{\;\;\;\;\infty ,\;\;\;\;\;\;\;\;\;\;\;\;\;\;{\rm{                                if\;}}\tilde \xi  \ge 1,}
\end{array}} \right.
\end{align}
where ${g_k} = \Gamma ( {1 - k\tilde \xi })$, $\tilde \xi  = 2\xi $, ${{\tilde e}_N}$ and ${{\tilde d}_N}$ are defined as
\begin{align}
{{\tilde d}_N} &\approx \ln \left( {1 + \bar \gamma {{\tilde b}_N}{{\tilde b}_N}} \right),\label{48A}\\
{{\tilde e}_N} &\approx \ln \left( {1 + \bar \gamma {{\left( {{{\tilde b}_N} + {{\tilde a}_N}} \right)}^2}} \right) - {{\tilde d}_N}.\label{49A}
\end{align}
\end{theorem}

\emph{Proof:} To start with, we let $\left\{ {{X_1}, \dots ,{X_N}} \right\}$, $\left\{ {{Y_1}, \dots ,{Y_N}} \right\}$, and $\left\{ {{Z_1},\dots ,{Z_N}} \right\}$ be i.i.d.~RV sequences, where ${Y_n} = {\left( {{X_n}} \right)^2}$ and ${Z_n} = \ln \left( {1 + \bar \gamma {Y_n}} \right)$ for $\forall n \in \left\{ {1, \dots ,N} \right\}$. Denote the maximum of the RV sequences as $M_N^X = \max \left\{ {{X_1},{X_2}, \dots ,{X_N}} \right\}$, $M_N^Y = \max \left\{ {{Y_1},{Y_2}, \dots ,{Y_N}} \right\}$, and $M_N^Z = \max \left\{ {{Z_1},{Z_2}, \dots ,{Z_N}} \right\}$, respectively. If ${{\left( {M_N^X - b_n^X} \right)} \mathord{\left/ {\vphantom {{\left( {M_N^X - b_n^X} \right)} {a_n^X}}} \right. \kern-\nulldelimiterspace} {a_n^X}}$ converges to a GEV distribution, then ${{\left( {M_N^Y - b_n^Y} \right)} \mathord{\left/ {\vphantom {{\left( {M_N^Y - b_n^Y} \right)} {a_n^Y}}} \right. \kern-\nulldelimiterspace} {a_n^Y}}$ and ${{\left( {M_N^Z - b_n^Z} \right)} \mathord{\left/ {\vphantom {{\left( {M_N^Z - b_n^Z} \right)} {a_n^Z}}} \right. \kern-\nulldelimiterspace} {a_n^Z}}$ also converge to the GEV distribution. The relationship between the normalizing parameters is given by $b_n^Y = b_n^Xb_n^X $, $a_n^Y = 2b_n^Xa_n^X + a_n^Xa_n^X$, $b_n^Z = \ln \left( {1 + \bar \gamma b_n^Y} \right)$ and $a_n^Z = \ln \left( {1 + \bar \gamma \left( {b_n^Y + a_n^Y} \right)} \right) - b_n^Z$. The relationship between the shape parameters is ${\xi _Y} = 2{\xi _X}$ and ${\xi _Z} = 0$ when ${\xi _X} > 0$; ${\xi _Y} = {\xi _Z} = 0$ when ${\xi _X} = 0$; ${\xi _Y} = {\xi _Z} = {\xi _X}$ when ${\xi _X} < 0$.
However, since the number of RVs $N$ is finite, the maximum samples may not sufficiently cover the tail, leading to a deviation of the actual shape parameter relationship from that in the asymptotic case. In the i.i.d. case, ${\xi _Z} \approx 2{\xi _X}$ with error $O\left( {{1 \mathord{\left/
 {\vphantom {1 {\ln N}}} \right.
 \kern-\nulldelimiterspace} {\ln N}}} \right)$ holds for finite $N$ through a penultimate extreme-value analysis combined with a second-order delta-method expansion.
Here, we fit the distribution of ${\left| {{h_{{\rm{FAS}}}}} \right|}$ using the GEV distribution, i.e., $\left| {{h_{{\rm{FAS}}}}} \right| \sim {\rm{GEV}}\{ {{{\tilde b}_N},{{\tilde a}_N},\xi } \}$, and thus the distribution of $C = \ln ( {1 + \bar \gamma {{\left| {{h_{{\rm{FAS}}}}} \right|}^2}} )$ also can be fitted by the GEV distribution, i.e., $C \sim {\rm{GEV}}\{ {{{\tilde e}_N},{{\tilde d}_N},\tilde \xi } \}$. From the normalizing parameters of $\left| {{h_{{\rm{FAS}}}}} \right|$ as shown in \eqref{23A}--\eqref{25A}, we derive ${{\tilde d}_N}$ and ${{\tilde e}_N}$ of $C$, respectively, as \eqref{48A} and \eqref{49A}.
Although spatial correlation exists among the $N$ RVs, the conclusion ${\xi _{Z,N}}\approx2{\xi _{X,N}}$ in the i.i.d. case for finite $N$ supports the use of the ``rule-of-thumb" approximation ${\tilde\xi}\approx2\xi$ to compute the shape parameter of the GEV distribution for fitting the distribution of $C$.

Until now, the normalizing parameters ${{\tilde d}_N}$ and ${{\tilde e}_N}$ and the shape parameter $\tilde \xi$ of the GEV distribution for fitting the distribution of $C$, can be found from ${{\tilde a}_N}$, ${{\tilde b}_N}$, and $\xi$. Therefore, ${\bar C}=\mathbb{E}\left\{ {C} \right\}$ can be approximated as \eqref{47A}.\hfill {$\blacksquare $}

\subsection{Simulation Results}
\subsubsection*{The accuracy of the fitted GEV distribution}
The results in Fig.~\ref{fig:cdf-gev} are provided to illustrate the CDFs of ${\left| {{h_{{\rm{FAS}}}}} \right|}$ obtained from the fitted GEV distribution under different $N$ and $W$. The black boxes in Fig. 6 indicate that the values in this region are shown in logarithmic scale in the small figures on the right. Compared to the Gumbel distribution, more accurate CDFs are obtained using the GEV distribution, especially in the extremely low-probability and high-probability regions. To verify its superiority in fitting accuracy, Fig.~\ref{fig:QQ-gev} provides the Q-Q plot of the empirical distribution versus both the fitted Gumbel and GEV distributions. It is evident that the GEV distribution provides an almost perfect fit to the exact channel distribution, outperforming the Gumbel distribution. Although minor deviations appear in the high-quantile region, they have a negligible effect on both the approximate OP and EC.

\subsubsection*{The accuracy of approximate OP}
\begin{figure}
\centering
\includegraphics[width=0.84\columnwidth]{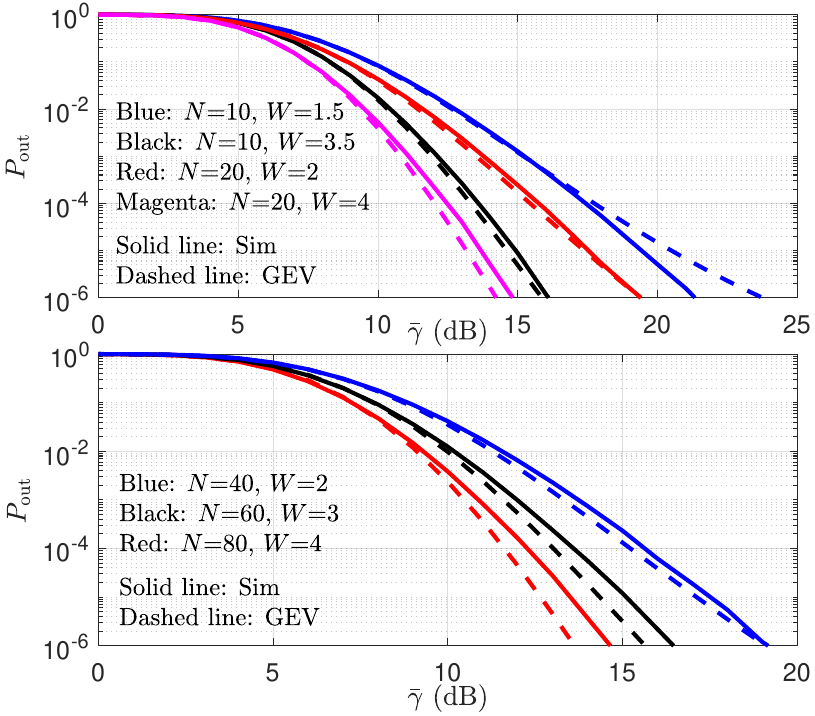}\\
\caption{The OPs of FAS obtained from Monte-Carlo simulations and the fitted GEV distribution versus the transmit SNR $\bar \gamma$ for selected values of $W$ and $N$.}\label{fig:op-gev}
\end{figure}

\begin{figure}
\centering
\includegraphics[width=0.83\columnwidth]{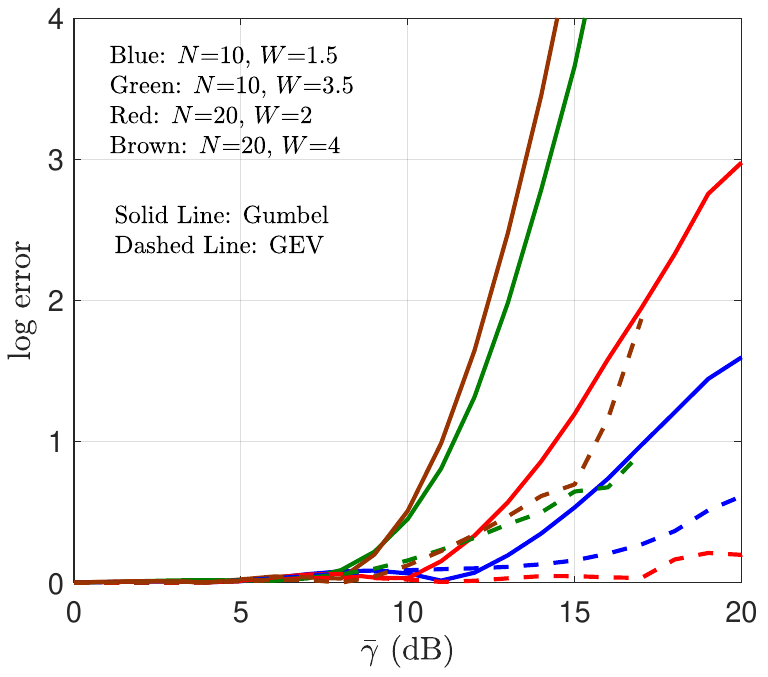}\\
\caption{The log-error between OPs obtained from Monte-Carlo simulations and those obtained from the fitted Gumbel and GEV distributions versus the transmit SNR $\bar \gamma$ under different $N$ and $W$.}\label{fig:op-error}
\end{figure}

We now provide the results in Fig.~\ref{fig:op-gev} for the approximate OPs of FAS versus the transmit SNR ${\bar \gamma }$ for different $N$ and $W$, with ${\gamma _{{\rm{th}}}} = 10$ dB. The results indicate that a satisfactory accuracy of the OP can be achieved using the GEV distribution fitting even in the low-OP region. To quantify the fitting error of OPs obtained from the EVDs, including the Gumbel and GEV distributions, Fig.~\ref{fig:op-error} shows the fitting log-error\footnote{The log-error, defined as $\left| {{{\log }_{10}}\left( {{P_{{\rm{sim}}}}} \right) - {{\log }_{10}}\left( {{P_{{\rm{fit}}}}} \right)} \right|$, is used because the OP is most relevant at very small values, where an absolute error would fail to reflect meaningful differences. The logarithmic scale highlights these discrepancies for a clearer comparison.} between the OPs obtained from the Monte-Carlo simulations and those obtained from the fitted Gumbel and GEV distributions. It can be observed that compared to the Gumbel distribution, the GEV distribution fitting significantly reduces the error, especially in the high transmit SNR region.

\subsubsection*{The accuracy of approximate EC}
\begin{figure}
\centering
\includegraphics[width=0.9\columnwidth]{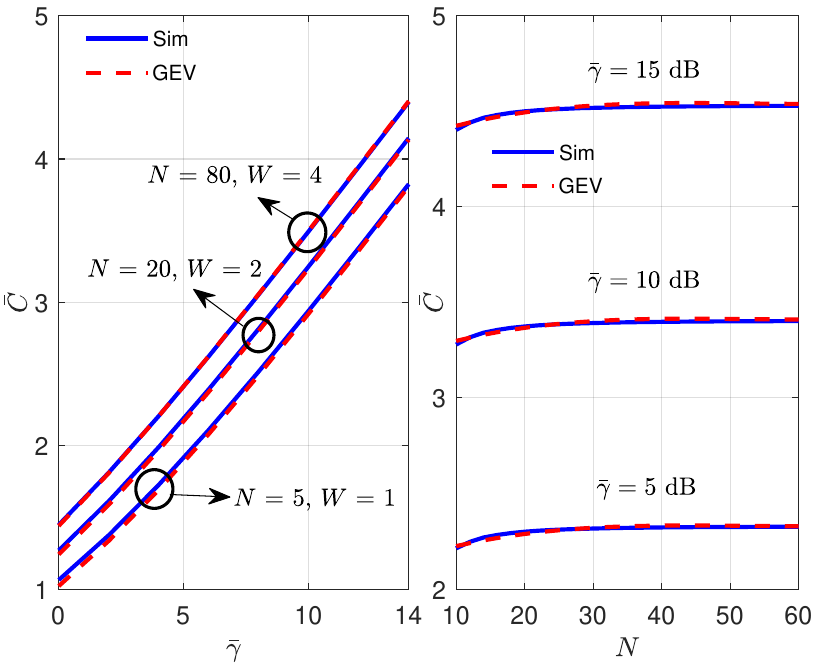}\\
\caption{The ECs of FAS obtained from the Monte-Carlo simulations and the fitted GEV distribution versus the transmit SNR $\bar \gamma$ and the
number of ports $N$, respectively.}\label{fig:ec-gev}
\end{figure}

\begin{figure}
\centering
\includegraphics[width=0.9\columnwidth]{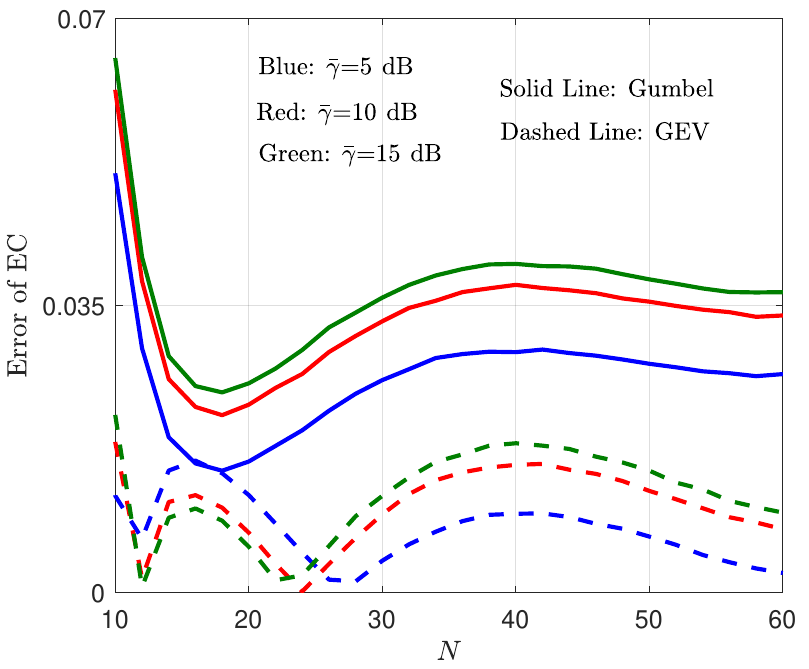}\\
\caption{The absolute error between ECs obtained from the Monte-Carlo simulations and those obtained from the fitted Gumbel and GEV distributions versus $N$, under different transmit SNR $\bar \gamma$.}\label{fig:ec-error}
\end{figure}

Fig.~\ref{fig:ec-gev} illustrates the approximate ECs of FAS versus the transmit SNR ${\bar \gamma }$ and the number of ports $N$, respectively. More accurate ECs are obtained using the fitted GEV distribution compared to the fitted Gumbel distribution. To illustrate this more intuitively, Fig.~\ref{fig:ec-error} presents the absolute error of the ECs obtained from the Monte-Carlo simulations and the fitted EVDs, including the fitted Gumbel and GEV distributions. Since the EC does not take very small values, the absolute error, defined as $\left| {{{\bar C}_{{\rm{sim}}}} - {{\bar C}_{{\rm{fit}}}}} \right|$, is sufficient to illustrate the fitting accuracy. We can see that the GEV distribution provides a superior fitting accuracy of the EC compared to the Gumbel distribution. The advantages of the GEV distribution lie in the introduction of a shape parameter, which allows flexible adaptation among the three types of EVD, and in its ability to better capture skewness and kurtosis, thereby more accurately reflecting the characteristics of the FAS channel distribution when $N$ is finite or strong correlation exists.

\subsection{Comparison of the GEV-Based Approach with Existing Methods}
To further verify the superiority of the proposed GEV distribution fitting method, we compare the performance evaluation results using the fitted GEV distribution with that in existing research in terms of the approximate OPs and the elapsed times for computing OPs, as listed in TABLE \ref{tab:time}.

\begin{figure}
\centering
\includegraphics[width=0.87\columnwidth]{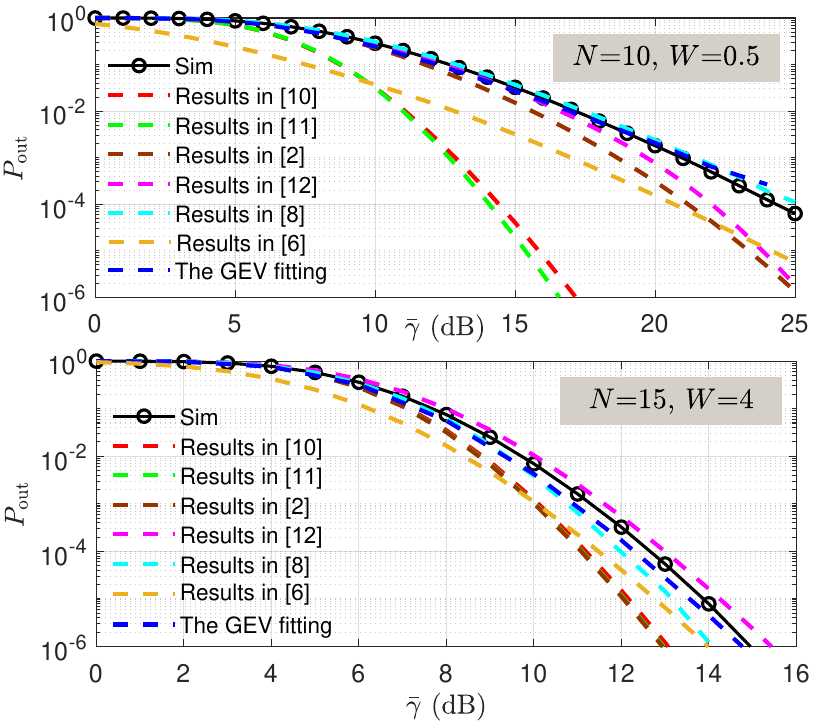}\\
\caption{The OPs of FAS obtained from different methods versus the transmit SNR $\bar \gamma$, under different $N$ and $W$.}\label{fig:op-compare-vs-snr}
\end{figure}

Fig.~\ref{fig:op-compare-vs-snr} presents the OPs obtained from different methods under two sets of parameters $N$ and $W$, with ${\gamma _{{\rm{th}}}} = 10$ dB. The Monte-Carlo simulations (marked as ``Sim") are used as a benchmark to evaluate the accuracy of the obtained OPs using various methods. Several observations can be made. First, our proposed GEV distribution fitting method using the obtained parameter expressions (marked as ``The GEV fitting") yields satisfactory accuracy of the OPs under both parameter settings. Second, the reference correlation model \cite{9264694} and the equally correlated model \cite{wong2022closed} significantly underestimate the OPs of FAS. Third, the two-stage approximation method under the fully correlated model \cite{10103838} provides reasonably accurate OP for $N=10,W=0.5$, but underestimates the OP for $N=15,W=4$. Fourth, the block-diagonal correlation model \cite{10623405} gives accurate OP for $N=15,W=4$ but this is no longer the case for $N=10,W=0.5$. Additionally, the Gaussian copula method \cite{10678877} has high accuracy for $N=10,W=0.5$, but shows significant deviations from the Monte-Carlo results for the case $N=15,W=4$, confirming its inaccuracy in sparse port deployments as reported in \cite{11023237,11230883}. For the FAS channel approximation in \cite{10924151}, although this approach offers extremely low computational complexity, its accuracy degrades noticeably in dense deployment.
As for the finite series expression \cite{10130117} and matrix approximation \cite{11023237} methods, they produce accurate OPs for very small port numbers (i.e., $N=2,3$), but their computational time increases dramatically for larger $N$, and therefore their OP results are not shown in Fig.~\ref{fig:op-compare-vs-snr}. From the above observations, we can conclude that existing methods only achieve reasonable accuracy of OP under specific parameter settings, whereas our proposed GEV distribution fitting consistently provides accurate OPs.

Moreover, we analyze the computational complexity of each method. TABLE \ref{tab:time} summarizes the runtime of various methods under the two system settings used in Fig.~\ref{fig:op-compare-vs-snr}. It can be seen that our proposed method has the shortest runtime among all of the methods. Although the reference correlation model \cite{9264694}, the equally correlated model \cite{wong2022closed}, and the block-diagonal correlation model \cite{10623405} also have relatively short runtimes, the accuracy of OP is insufficient. Notably, the Gaussian copula method \cite{10678877} and the two-stage approximation method under the fully correlated model \cite{10103838} require considerably more time, and the finite series expression \cite{10130117} and the matrix approximation \cite{11023237} methods are even more time-consuming. Therefore, our proposed GEV distribution fitting achieves a favorable balance between accuracy and computational complexity, providing a fast, precise method for evaluating the FAS performance.

\begin{figure}
\centering
\includegraphics[width=0.9\columnwidth]{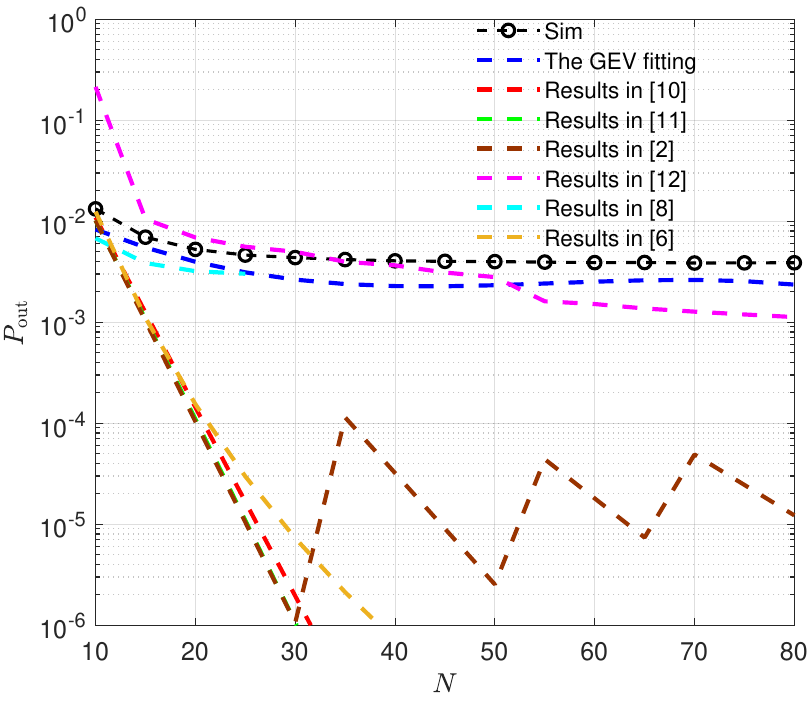}\\
\caption{The OPs of FAS obtained from different methods versus the number of ports $N$ with $W=4$.}\label{fig:op-compare-vs-N}
\end{figure}

Furthermore, Fig.~\ref{fig:op-compare-vs-N} shows the OPs obtained from different methods versus the number of ports $N$. It can be observed that the accuracy of our proposed method remains consistently high as $N$ varies. The reference correlation \cite{9264694}, the equally correlated \cite{wong2022closed}, and the block-diagonal correlation \cite{10623405} models consistently underestimate the OP. Moreover, the two-stage approximation method \cite{10103838} under fully correlated model underestimates the OPs for large port numbers. For the Gaussian copula method \cite{10678877}, only results for small $N$ are shown, as the computation becomes infeasible for larger $N$ due to memory limitations, highlighting a key drawback of this method.

\begin{table}[t]
\centering
\caption{The elapsed time for computing OPs in Fig.~\ref{fig:op-compare-vs-snr}}\label{tab:time}
\begin{tabular}{>{\centering\arraybackslash}m{4cm}|c|c}
\hline
\multirow{2}{*}{Various methods} & \multicolumn{2}{c}{Time for computing OPs} \\
\cline{2-3}
 & $\left\{\begin{aligned}
 N&=10\\
 W&=0.5
 \end{aligned}\right.$
  & $\left\{\begin{aligned}
 N&=15\\
 W&=4
 \end{aligned}\right.$\\
\hline\hline
The GEV distribution fitting & $0.031$ s & $0.020$ s \\
\hline
The reference correlation model \cite{9264694} & $0.268$ s & $0.334$ s \\
\hline
The equally correlated model \cite{wong2022closed} & $0.119$ s & $0.135$ s \\
\hline
The fully correlated model \cite{10103838} & $1.404$ s & $>1$ min \\
\hline
The block-diagonal correlation model \cite{10623405} & $0.222$ s & $0.547$ s \\
\hline
The Gaussian copula method \cite{10678877} & $54.086$ s & $58.088$ s \\
\hline
The finite series expression \cite{10130117} & $>1$ min & $>1$ min \\
\hline
The matrix approximation \cite{11023237} & $>1$ min & $>1$ min \\
\hline
{The\! FAS\! channel\! approximation}\cite{10924151} & {$0.011$ s} & {$0.0109$ s} \\
\hline
\end{tabular}
\end{table}

\section{Conclusions}\label{sec:conclude}
In this paper, we proposed a novel EVD-based performance evaluation framework for FAS that achieves accurate yet computationally efficient performance evaluation. The distribution of $\left|h_{\text{FAS}}\right|$ was fitted using both the Gumbel and the more flexible GEV distributions. The corresponding parameters were expressed as functions of the number of ports $N$ and the antenna size $W$, based on which closed-form approximations of OP and EC were derived. Three main conclusions can be drawn.
First, the Gumbel distribution provides a generally accurate fit for the FAS channel distribution, although slight deviations occur in the high- and ultra-low-probability regions. As a result, a simple and sufficiently accurate expression for EC is obtained, whereas slight deviations for OP occur in the low-probability region.
Second, the GEV distribution offers a more accurate characterization of the FAS channel, yielding improved approximations of OP and EC. Third, compared with existing methods, the proposed GEV-based framework achieves accurate OP and EC estimations with significantly lower computational complexity.
The fitted closed-form FAS channel distribution not only enables accurate and efficient performance evaluation, but also facilitates performance analysis of more complex systems, such as interference-limited FAS or fluid antenna-aided backscatter communication systems. This constitutes an important direction for future work.

Despite this, several practical considerations and limitations remain open topics.
Although a satisfactory accuracy of OP
can be achieved using the fitted GEV distribution, slight deviations are present in the ultra-low-probability region, which is critical for ultra-reliable communications. Therefore, when the primary objective of channel modeling
is outage performance evaluation, one may further improve accuracy by adopting a tail-aware EVD or
fitting criteria that place greater emphasis on the left tail, which provides an interesting direction for future
research.
Besides, this paper only considers a straight-line FAS and models its channel using the distribution fitting
technique to facilitate performance evaluation.
In practical implementations, however, fluid antennas may operate in 2D or 3D geometries
(e.g., inside a capsule, ring, or volumetric structure), where the induced spatial correlation structure would change. How to employ distribution fitting techniques to accurately model such channels remains an open issue.
Moreover, the idealized channel statistics and correlation matrix are assumed to be perfectly known, while channel estimation errors \cite{10751774, 11180048} and outdated channel state information in practical systems may affect the accuracy of the fitted distributions, which also warrants further investigation.
Finally, although the proposed framework is independent of the specific hardware realization of fluid antennas, achieving fast and low-complexity port switching is a key practical requirement for fluid antenna implementations.

\ifCLASSOPTIONcaptionsoff
  \newpage
\fi
\bibliographystyle{IEEEtran}

\end{document}